\documentclass[a4paper,11pt]{article}
\pdfoutput=1
\usepackage{amssymb,amsmath,amsfonts,makeidx,placeins,pbox,multirow}
\usepackage{graphicx,rotate,subcaption,color,slashed,cite,caption,epstopdf,verbatim}
\usepackage{longtable,tabu}
\usepackage{array}
\usepackage{color}
\usepackage[table]{xcolor}
\usepackage{amsmath}
\usepackage{mathtools}
\usepackage{amssymb}
\usepackage{hyperref}
\usepackage[normalem]{ulem}
\newcolumntype{P}[1]{>{\centering\arraybackslash}p{#1}}

\def\lapp{\mathrel{\rlap{\raise.5ex\hbox{$<$}}
                    {\lower.5ex\hbox{$\sim$}}}}
\def\gapp{\mathrel{\rlap{\raise.5ex\hbox{$>$}}
                  {\lower.5ex\hbox{$\sim$}}}}

\usepackage{color}
\long\def\/*#1*/{}
\usepackage{color}
 \usepackage[normalem]{ulem}
\definecolor{darkgreen}{cmyk}{1,0,1,0.4}
\definecolor{darkred}{cmyk}{0,1,1,0.4}
\definecolor{rosso}{cmyk}{0,1,1,0.4}
\definecolor{rossos}{cmyk}{0,1,1,0.55}
\definecolor{rossoc}{cmyk}{0,1,1,0.2}
\definecolor{blu}{cmyk}{1,1,0,0.3}
\definecolor{blus}{cmyk}{1,1,0,0.6}
\definecolor{bluc}{cmyk}{1,1,0,0.1}
\definecolor{verde}{cmyk}{0.92,0,0.59,0.25}
\definecolor{verdec}{cmyk}{0.92,0,0.59,0.15}
\definecolor{verdes}{cmyk}{0.92,0,0.59,0.4}
\definecolor{grigio}{cmyk}{0,0,0,0.07}
\definecolor{rosa}{cmyk}{0,0.1,0.1,0.02}
\definecolor{rosino}{cmyk}{0,0.05,0.05,0.02}
\definecolor{rosas}{cmyk}{0,0.3,0.25,0.05}
\definecolor{celeste}{cmyk}{0.1,0,0,0.02}
\definecolor{giallino}{cmyk}{0,0,0.4,0.02}
\definecolor{rosso}{cmyk}{0,1,1,0.4}
\definecolor{rossos}{cmyk}{0,1,1,0.55}
\definecolor{rossoc}{cmyk}{0,1,1,0.2}
\definecolor{blu}{cmyk}{1,1,0,0.3}
\definecolor{bluc}{cmyk}{1,1,0,0.1}
\definecolor{blucc}{cmyk}{0.7,0.5,0,0}
\definecolor{viola}{cmyk}{0,1,0,0.6}
\definecolor{viola2}{cmyk}{0,1,0.2,0.6}
\definecolor{verde}{cmyk}{0.92,0,0.59,0.25}
\definecolor{verdec}{cmyk}{0.92,0,0.59,0.15}
\definecolor{verdes}{cmyk}{0.92,0,0.59,0.4}
\definecolor{verdino}{cmyk}{0.12,0,0.09,0.05}
\definecolor{giallo}{cmyk}{0,0,1,0}
\definecolor{gialloverde}{cmyk}{0.44,0,0.74,0}

\evensidemargin=\oddsidemargin
\def\bar {\overline}

\def\bea {\begin{eqnarray}}
\def\eea {\end{eqnarray}}

\def\beq{\begin{equation}}
\def\eeq{\end{equation}}
\def\barr{\begin{array}}
\def\earr{\end{array}}
\def\dis{\displaystyle}

\newcommand{\nn}{\nonumber \\}

\def\beq{\begin{equation}}
\def\eeq{\end{equation}}

\newcommand{\nc}{\newcommand}

\newcommand{\C}{\mathcal{C}}
\newcommand{\Op}{{\cal O}}
\nc{\hi}{H}
\nc{\hit}{\widetilde{H}}
\nc{\hij }{\mbox{${\hi^\dag i\,\raisebox{2mm}{\boldmath ${}^\leftrightarrow$}\hspace{-4mm} D_\mu\,\hi}$}}
\nc{\hijt}{\mbox{${\hi^\dag i\,\raisebox{2mm}{\boldmath ${}^\leftrightarrow$}\hspace{-4mm} D_\mu^{\,a}\,\hi}$}}

\def\nn{\nonumber}
\def\gev{\rm GeV}
\def\tev{\rm TeV}

\def\gev{\,\ensuremath{\mathrm{Ge\kern -0.1em V}}}
\def\tev{\,\ensuremath{\mathrm{Te\kern -0.1em V}}}

\DeclareUnicodeCharacter{2212}{-}
\UseRawInputEncoding
\begin{document}
\begin{center}
{\Large {\bf Constraining SMEFT BSM scenarios with
 EWPO and $\Delta_{CKM}$}} \\
\vspace*{0.8cm} {\sf  ${^1}$Mathew Thomas Arun \footnote{thomas.mathewarun@gmail.com}, ${^2}$Kuldeep Deka\footnote{kuldeepdeka.physics@gmail.com}, ${^2}$Tripurari Srivastava\footnote{tripurarisri022@gmail.com}} \\
\vspace{10pt} {\small } {\em  ${^1}$School of Physics, Indian Institute of Science Education and Research Thiruvananthapuram, Vithura, Kerala, 695551, India\\

${^2}$Department of Physics and Astrophysics, University of Delhi, Delhi 110007, India\\}
\normalsize
\end{center}
\bigskip
\begin{abstract}
 
Precision observables are well known for constraining most of the Beyond Standard Model (BSM) scenarios tightly. We present here a simple and comprehensive fitting framework for various BSM scenarios to these observables. We start with the fit of $S$, $T$ and $V$ parameter and their correlations using the Electroweak Precision Observables (EWPO) including the recent $m_W$ measurement from CDF-II. Utilizing these observables, we also fit various New Physics (NP) scenarios consisting of different subsets of dimension-6 Standard Model Effective Field Theory (SMEFT) operators in the Warsaw basis out of a total of 10 appearing at tree level in EWPO. To further constrain these scenarios, we augment these observables with $\Delta_{CKM}$ measurement using 1-loop matching of the Low Energy Effective Field Theory (LEFT) to SMEFT operators at the Z-pole. We show that the inclusion of $\Delta_{CKM}$ constraint indeed results in stronger bounds on the SMEFT Wilson Coefficients. We also constrain the UV parameters of BSM extensions like Vectorlike leptons (VLL) and find out that such a minimal extension is in tension with the forward-backward asymmetry in $b$-sector ($A_b^{FB}$) and the recent measurement of $M_W$. In order to lift the two blind directions, which one encounters while fitting all the 10 SMEFT WCs at tree-level, we also include the LEP-II observables pertaining to the $WW$ production and present the results for the fits with and without $\Delta_{CKM}$ constraint.

\end{abstract}
\section{Introduction}
\label{sec:intro}
The discovery of the Higgs at the LHC in 2012 \cite{ATLAS:2012yve} completes the Standard Model of particle physics (SM). On the other hand, the absence of any heavy resonances in the Run 2 of the Large Hadron Collider (LHC) implies that any physics beyond the Standard Model (BSM) is likely to be associated with an intrinsic energy scale significantly higher than the electroweak scale. Interestingly, however, a few discrepancies have been seen in low energy experiments like the ($g-2$) of the muon \cite{Aoyama:2020ynm,Muong-2:2021ojo,Choudhury:2022iqz,Deka:2022ltk,Chakrabortty:2015zpm}, flavor anomalies~\cite{Du:2021zkq,Ban:2021tos,Darme:2021qzw,Bhattacharya:2021ggm}, the Cabibbo anomaly \cite{10.1093/ptep/ptac097} and the recent CDF anomaly in W boson mass~\cite{CDF:2022}. Given the elusiveness of New Physics at collider experiments, there is
a huge motivation to understand the nature of various anomalies in a model-independent Effective Field Theory formalism. In our study, a class of New Physics effects are included by extending SM with higher dimensional operators which are invariant under the SM gauge group. 
While the full set of such operators $-$ together constituting the Standard Model Effective Field Theory (SMEFT)~\cite{Lehman:2015coa,Henning:2015alf,Grzadkowski:2010es,Kennedy:1988sn,Buchmuller:1985jz,Weinberg:1979sa} $-$ is large, we concentrate on a subset that are likely to play leading roles in addressing the Cabibbo and CDF anomalies as well as the Electroweak Precision Observables (EWPO). Being well measured at LEP~\cite{ALEPH:2005ab,ALEPH:2004dmh,OPAL:2007ytu,L3:2004lwm,ALEPH:2013dgf} along with SLAC and Tevatron~\cite{ALEPH:2010aa}, the precise measurements of Electroweak Precision Observables (EWPO) strongly constrain the scales of these SMEFT operators.

In the last decade, studies on SMEFT have gained significant amount of attention. Many studies have been performed to analyze the SMEFT in the context of EWPO~\cite{Falkowski:2014tna,Henning:2014wua,Berthier:2015oma,Berthier:2015gja,Brivio:2017vri, Ellis:2018gqa, Dawson:2019clf, Dawson:2020oco,Ellis:2020unq,Ethier:2021bye}. At tree level in the Warsaw basis~\cite{Grzadkowski:2010es}, eight SMEFT WCs out of a total of ten contributing to the EWPO can be constrained~\cite{Brivio:2017bnu}. It has also been found that the inclusion of WW production data at LEP-II, allows for constraining all the ten WCs~\cite{Berthier:2016tkq,Brivio:2017bnu, Brivio:2017bnu}. Moreover, there have also been numerous studies on the global fit of EWPO in the context of SMEFT~\cite{Berthier:2016tkq, Ellis:2018gqa, deBlas:2022ofj, Ethier:2021bye}.
    
The recent measurement of $W$ boson mass~\cite{CDF:2022}, 
$M_W =80,433.5 \pm 9.4$ MeV, disagrees with the SM prediction, $M_W=80,360\pm 6$ MeV, at $7\sigma$ level. The new combined world average, calculated from the uncorrelated average of all $m_W$ measurements at Tevatron, LEP-II and the LHC, now lies at $M_W=80,413.3\pm 9$ MeV~\cite{deBlas:2022hdk}. Several studies have been conducted to address the impact of this new measurement on 
the Global EW fit in terms of the oblique parameters~\cite{Lu:2022bgw,Strumia:2022qkt,Asadi:2022xiy,Paul:2022dds} and also assuming the SMEFT framework ~\cite{deBlas:2022hdk,Fan:2022yly,Bagnaschi:2022whn,Gu:2022htv,Endo:2022kiw,Balkin:2022glu,Gupta:2022lrt,Cirigliano:2022qdm,Bagnaschi:2022whn}. Models such as Technicolor models, Extradimensions, Composite Higgs or Little Higgs scenarios, SU(2) Triplet Scalar Extensions~\cite{Kanemura:2022ahw}, Models with VLQ~\cite{Cao:2022mif}, Two Higgs Doublet Models~\cite{Lu:2022bgw} among many others~\cite{Athron:2022qpo,Yang:2022gvz,Yuan:2022cpw,Blennow:2022yfm,Strumia:2022qkt} have also been proposed to address this.

Apart from this, anomalies have been reported in low-energy measurements. These are studied using Low-energy Effective Field Theory (LEFT) description of the Standard Model. In particular, in this study, we have considered the beta decay and other semileptonic processes that contribute to the deviation in first row unitarity of the $CKM$ matrix ($|V_{ud}|^2 + |V_{us}|^2 - 1)$, neglecting the tiny $|V_{ub}|^2$ contributions) in the LEFT basis. Also known as $\Delta_{CKM}$ or Cabibbo anomaly, this quantity shows a discrepancy of 2$\sigma$. Since $\Delta_{CKM}$ and electro-weak precision tests crucially depend on the measured value of $G_F$, albeit at different scales, these measurements are connected and have to be taken together while constraining any New Physics. To that end, we match these LEFT operators with SMEFT at 1-loop at the $Z$ boson scale and obtain experimental constraints on the Wilson Coefficients (WCs) of the dimension-6 SMEFT operators. 

In this article, we perform an all-parameter fit which takes into account all the 10 WCs affecting the EWPO at tree-level, with the recent CDF-II $M_W$ mass measurement and $\Delta_{CKM}$ constraint. Along with these two observables, $A_b^{FB}$ and $A_l^{SLD}$ too play significant roles in these fits, since both have a long-standing discrepancy of more than 2$\sigma$ with respect to the SM predictions. However, we encounter two blind directions as the EWPO can only constrain 8 independent combinations of the WCs. Addition of $\Delta_{CKM}$ also does not help in lifting the blind directions. Rather, this forces us to use LEP-II data, where the presence of W-W production channel breaks the two blind directions, enabling us to constrain all the 10 WCs of our interest. In order to realise these fits, we have written our own code from the scratch in {\it Mathematica}, making it highly expandable and modular in design. To make it more versatile and user-friendly, we are also working on its easy integration with other publicly available packages for various SMEFT and New Physics calculations like CoDEx\cite{DasBakshi:2018vni}, SmeftFR \cite{Dedes:2023zws}, SMEFTsim\cite{Brivio:2020onw}, DsixTools\cite{Fuentes-Martin:2020zaz}, Rosetta \cite{Falkowski:2015wza}, FeynMG \cite{SevillanoMunoz:2022tfb}, FeynRules\cite{Alloul:2013bka}, FeynCalc\cite{Shtabovenko:2020gxv}, FeynArts\cite{Hahn:2000kx} {\it etc}, other LEFT-SMEFT matching tools \cite{Dekens:2019ept} and tools for computing Electroweak chiral amplitudes (Higgs EFT) \cite{Martinez-Martin:2022loy}.

In the SMEFT parametrisation, we also consider certain subsets of operators affecting the observables of our interest and perform the fit to constrain their WCs. Depending on the UV complete scenarios, typically only a subset of all the WCs contributes significantly to a set of well measured low energy observables. One such fit looks at the scenario where the operators common to the $M_W$ and $\Delta_{CKM}$ are the ones carrying the imprint of the New Physics. Another New Physics scenario considered here is the set of WCs which appear in the expression for $M_W$ at the tree-level.

We also perform a fit for the vector-like lepton models which are popular for their ability to address the discrepancy in muon $(g-2)$ and providing masses to the SM neutrinos through the seesaw mechanism. By matching these models to SMEFT at the tree-level, we perform the fit where the WCs now carry the imprints of couplings and the mass-scale of these new leptons. It also enables us to constrain the UV parameters of the model.

Moreover, we have used $S,~T$ and $V$~\cite{Peskin:1991sw,Maksymyk:1993zm,Burgess:1993vc} to parametrise the New Physics contribution to EWPO. With the assumption that there are no additional electroweak gauge bosons and that new physics coupling to light fermions are suppressed, $S$ and $T$ (along with $U$, contributions to which from majority of new physics scenarios are small and generated at SMEFT in dimension-8) parametrise the new physics effects in the vacuum polarisation diagrams of the electroweak gauge bosons. If the presence of New Physics also leads to changing of the weak coupling constant $G_F$, it can be parametrised by the inclusion of $V$ parameter along with $S$ and $T$~\cite{Csaki:2002gy}. Since the Cabibbo anomaly can only be explained by adjusting the $V$ parameter, this analysis is rendered all the more important.

This paper is organized as follows: In the next section, we revisit the contributions of dimension-6 operators to the shift in fermion couplings to gauge-bosons at leading order. In section~[\ref{sec:zpole}], we briefly discuss the near $Z$ pole precision observables (EWPO) and the observables arising due to $WW$ production (LEP-II). We then discuss the contributions of the SMEFT operators to muon and beta decay as obtained by matching with LEFT at one loop (in section)~[\ref{sec:LEFT}] and discuss the fitting framework in section~[\ref{sec:ewf}]. Further, we analyze various model independent frameworks, using $S,~T, ~V$ and the subsets of dimension-6 operators affecting the EWPO and $\Delta_{CKM}$. Finally, in section~[\ref{sec:outlook}] we summarize the results.

\section{SMEFT contributions to Electwoweak Fit}
\label{sec:basics}
The low-energy effects of massive BSM particles
can be approximated by integrating out the corresponding fields to obtain higher-dimensional interactions between the SM fields. In such an approach, the SM can be regarded as an effective theory whose known renormalizable interactions are supplemented by higher-order terms scaled by inverse powers of the BSM mass scale. The consequent effective Lagrangian can, then, be written in the form
\begin{equation}
{\cal L}_{eff}={\cal L}_{SM}+\sum_{d=5}^{\infty}\sum_{i=1}^n {\mathcal{C}_i^d\over \Lambda^{d-4}} O_i^d\, ,
\label{eq:lsmeft}
\end{equation}
where $d$ represents the dimension of the operator and $i$ runs over all the relevant set of operators at a particular dimension. The operators $O_i^d$ are all $SU(3)\times SU(2)_L\times U(1)_Y$ invariant with all of  the effects of the BSM  physics  residing in the WCs $\C_i^d$. 
In our work, we assume the WCs to be real to eliminate any new sources of CP violation and use the Warsaw basis \cite{Grzadkowski:2010es} to parametrise them. 

The EWPO of our primary interest are the $Z$ and $W$ pole observables and there are 10 six dimensional operators which contribute to these observables at the tree level. 
We collect these operators in Table \ref{tab:opdef}, where $\hi$ is the $SU(2)_L$  Higgs doublet,  $\tau^a$  are the Pauli matrices, 
$D_\mu=\partial_\mu +ig_s
T^AG_\mu^A +ig_2 {\tau^a\over 2}W_\mu^a+ig_1Y B_\mu$, $q$ is the $SU(2)_L$ quark-doublet with
$q^T=(u_L,d_L)$, $l$ is the $SU(2)_L$ lepton-doublet with $l^T=(\nu_L, e_L)$, $W_{\mu\nu}^ a$ is the $SU(2)_L$ field strength with  $W_{\mu\nu}^ a=
\partial_{\mu}W^a_{\nu} - \partial_{\nu}W^a_{\mu}-g_2\epsilon^{abc}W^b_{\mu}W^c_{\nu}$ and $B_{\mu\nu}$ is the $U(1)_Y$ field strength with  $B_{\mu\nu}=
\partial_{\mu}B_{\nu} - \partial_{\nu}W_{\mu}$. The other definitions include:
$\hij=i\hi^\dagger(D_{\mu} \hi)
-i(D_{\mu}\hi)^\dagger \hi$,
and
$\hijt=i\hi^\dagger \tau ^a D_{\mu} \hi
-i(D_{\mu}\hi)^\dagger \tau^a\hi$. 
 
\begin{table}[t] 
\begin{small}
\centering
\renewcommand{\arraystretch}{1.5}
\begin{tabular}{|c|c||c|c||c|c|} 
\hline
$\Op_{\hi D}$   & $\left(\hi^\dag D^\mu\hi\right)^* \left(\hi^\dag D_\mu\hi\right)$  &    
  ${\Op}_{\hi W B}$ 
 &$ (\hi^\dag \tau^a \hi)\, W^a_{\mu\nu} B^{\mu\nu}$  
 \\
 \hline
${\Op_{ll}}$                   & $(\bar l \gamma_\mu l)(\bar l \gamma^\mu l)$
 &
 
   ${\Op}_{\hi e}$  &   $(\hij) (\overline {e}\gamma^\mu e)$  
\\
\hline   
    ${\Op}_{\hi u}$ & $(\hij) (\overline {u}\gamma^\mu u)$ &
  ${\Op}_{\hi d}$
       & $(\hij) (\overline {d}\gamma^\mu d)$
  \\ \hline 
             ${\Op}_{\hi q_3}$ & $(\hijt)(\bar q \tau^a \gamma^\mu q)$  &${\Op}_{\hi q_1}$
      &$(\hij)(\bar q \gamma^\mu q)$  
\\
\hline      
 ${\Op_{\hi l_3}}$      & $(\hijt)(\bar l \tau^a \gamma^\mu l)$ &
${\Op}_{\hi l_1}$
      &  $(\hij)(\bar l \gamma^\mu l)$ 

\\
\hline 
\end{tabular}
\caption{Dimension-6 operators in Warsaw basis contributing to the EWPO at tree level. \label{tab:opdef}}
\end{small}
\end{table}

In order to get the theoretical predictions from the Electroweak precision Data pertaining to the pole observables, we first fix our choice of input parameters\footnote{The choice of this particular combination is inspired from the higher experimental precision in their determination as compared to the other parameters in the SM}, namely,
the fine structure constant $\hat{\alpha}_{e}$ from the low energy limit of electron Compton scattering, the Fermi constant in muon decays $\hat{G}_F$ and the measured $Z$ mass ($\hat{m}_Z$). At the tree level, one can then define the effective (measurable) mixing angle
\bea\label{sinequation}
s_W^2 = \frac{1}{2} - \frac{1}{2}\sqrt{1 - \frac{4 \, \pi \hat{\alpha}_{e}}{\sqrt{2} \, \hat{G}_F \, \hat{m}_Z^2}}.
\eea
The value of the $\rm SU(2)_L$ gauge coupling can be taken as:
\bea
 \hat{g}_2 \, s_W = 2 \, \sqrt{\pi} \, \hat{\alpha}_{e}^{1/2}.
\eea
The effective measured vacuum expectation value (vev) in the SM can be defined as $\hat{v}^2 = 1/\sqrt{2} \, \hat{G}_F$.

Once we allow for these new SMEFT operators, the gauge sector gets modified. To have canonical kinetic term, we must then effect a redefinition of fields and couplings. The redefined quantities are \cite{Dedes:2017zog}:
\begin{eqnarray}
{\overline W}_\mu^a & \equiv & (1-\C_{\hi W} v^2/\Lambda^2)W_\mu^a
\nonumber \\
{\overline B}_\mu & \equiv & (1-\C_{\hi B}v^2/\Lambda^2)B_\mu
\nonumber \\
{\overline g}_2 & \equiv &(1+\C_{\hi W} v^2/\Lambda^2)g_2
\nonumber \\
{\overline g}_1 & \equiv&  (1+\C_{\hi B}v^2/\Lambda^2)g_1\, ,
\end{eqnarray}
such  that ${\overline W}_\mu {\overline g}_2 \approx W_\mu g_2$ and ${\overline B}_\mu {\overline g}_1 \approx B_\mu g_1$.
The masses of the W and Z fields to ${\cal {O}}\biggl({\Lambda^{-2}}\biggr)$ are
  \cite{Dedes:2017zog},
\bea
M_W^2&=&\frac{{\overline g}_2^2 v^2}4,\nn\\
M_Z^2&=&\frac{({\overline g}_1^2+{\overline g}_2^2) v^2}4+\frac{v^4}{\Lambda^2}\left(\frac18 ({\overline g}_1^2+{\overline g}_2^2) \C_{\hi D}+\frac12 {\overline g}_1{\overline g}_2\C_{\hi WB} \right).
\eea
The induced change to the effective 4-fermion operator governing the decay of the muon alters the relation between the
vev, $v$, and the Fermi constant $\hat{G}_F$ to 
\begin{eqnarray}
\hat{G}_F
\equiv \frac1{\sqrt{2} v^2}-\frac1{\sqrt{2}\Lambda^2}\C_{ll}+{\sqrt{2}\over \Lambda^2}\C_{\hi l_3}\, .
\label{eq:gdef}
\end{eqnarray}
Using dimensionful WCs $C_i = \C_i/\Lambda^2$ (in $TeV^{-2}$) for the subsequent parts, the shift in $M_Z^2$ can then be written as:
\bea\label{deltaMz}
\delta M_Z^2 \equiv  \frac{1}{2 \, \sqrt{2}} \, \frac{\hat{m}_Z^2}{\hat{G}_F} C_{HD} + \frac{2^{1/4} \sqrt{\pi} \, \sqrt{\hat{\alpha_e}} \, \hat{m}_Z}{\hat{G}_F^{3/2}} C_{HWB}.
\eea
where $\hat{m}_Z$ is the input parameter 
The kinetic mixing introduced by the operator $\mathcal{O}_{HWB}$ leads to a shift in the usual $(s_W \equiv \sin \theta_W)$ mixing angle of the SM given by
\bea
\delta s_W^2 = -v^2 \frac{s_W^{} c_W^{}}{c_W^2-s_W^2}\left[2 s_W^{} c_W^{}\left(\delta v + \frac14 C_{\hi D}^{}\right) + C_{\hi WB}^{}\right].
\eea

We relate the Lagrangian parameters $\bar{g}_2,\bar{g}_1$ to the input parameters at tree level via
\bea
\bar{g}_1^2  + \bar{g}_2^2  = 4 \, \sqrt{2} \, \hat{G}_F \, \hat{m}_Z^2 \left(1 - \sqrt{2} \,\delta G_F - \frac{\delta M_Z^2}{\hat{m}_Z^2} \right), \\
\bar{g}_2^2 = \frac{4 \, \pi \, \hat{\alpha_e}}{s^2_W} \left[1 +  \frac{\delta s_W^2}{s^2_W} + \frac{c_W}{ s_W} \frac{1}{\sqrt{2} \, \hat{G}_F} \, C_{HWB} \right].
\eea

Expressing $\bar{M}_W^2$ in terms of the inputs parameters we get
\bea
\bar{M}_W^2 = \hat{m}_W^2 \left( 1 + \frac{\delta s_W^2}{s_W^2}+\frac{c_W}{s_W \sqrt{2} \hat{G}_F}C_{HWB} + \sqrt{2} \delta G_F \right) = \hat{m}_W^2 - \delta M_W^2,
\eea
where $\delta M_W^2= -\hat{m}_W^2 \left(\frac{\delta s_W^2}{s_W^2}+\frac{c_W}{s_W \sqrt{2} \hat{G}_F}C_{HWB} + \sqrt{2} \delta G_F\right)$
and $\hat{m}_W^2 = c_W^2 \hat{m}_Z^2 $.

The effective axial and vector couplings of the $Z$ boson are now defined as
\bea
\mathcal{L}_{Z,eff}  =  g_{Z,eff}  \,   \left(J_\mu^{Z \ell} Z^\mu + J_\mu^{Z \nu} Z^\mu + J_\mu^{Z u} Z^\mu +  J_\mu^{Z d} Z^\mu \right),
\eea
where $g_{Z,eff} = - \, 2^{5/4} \, \sqrt{\hat{G}_F} \, \hat{m}_Z$, and $(J_\mu^{Z x})^{ij} = \bar{\psi}_i \, \gamma_\mu \left[(\bar{g}^{x}_V)_{eff}^{ij}- (\bar{g}^{x}_A)_{eff}^{ij} \, \gamma_5 \right] \psi_j$ for $\psi = \{u,d,\ell,\nu \}$.
In general, these currents are matrices in flavour space. When we restrict our attention to the case of a minimal flavour violating (MFV) scenario, $(J_\mu^{Z x})_{ij} \simeq (J_\mu^{Z x}) \delta_{ij}$. The effective axial and vector couplings are modified from the SM values by a shift defined as
\bea
\delta (g^{x}_{V,A})_{ij} = (\bar{g}^{x}_{V,A})^{eff}_{ij} - (g^{x}_{V,A})^{SM}_{ij},
\eea
 
The tree level couplings are usual SM relations,
\begin{eqnarray}
g_R^{Zf}&=&-s_W^2 Q_f\quad{\rm and}\quad g_L^{Zf}=T_3^f -s_W^2 Q_f
\end{eqnarray}
with $T_3^f=\pm \displaystyle \frac{1}{2}$. 
The full SMEFT contributions to the effective couplings are shown in Table~\ref{tab:coupling-shifts}.

\begin{table}
\centering
\begin{tabular}{||c||c||}
\hline\hline
\rowcolor{celeste} $\delta (g^{\ell}_V)$&$\delta \bar{g}_Z \, (g^{\ell}_{V})^{SM} - \frac{1}{4 \sqrt{2} \hat{G}_F} \left(C_{He} + C_{Hl_1} + C_{Hl_3} \right) - \delta s_W^2, $\\
\hline
\rowcolor{verdino} $\delta(g^{\ell}_A)$&$\delta \bar{g}_Z \, (g^{\ell}_{A})^{SM}_{pr} + \frac{1}{4 \, \sqrt{2} \, \hat{G}_F}
\left( C_{He} - C_{Hl_1} - C_{Hl_3} \right),  $\\
\hline
\rowcolor{rosa} $\delta (g^{\nu}_V)$&$\delta \bar{g}_Z \, (g^{\nu}_{V})^{SM}_{pr} - \frac{1}{4 \, \sqrt{2} \, \hat{G}_F} \left( C_{Hl_1} - C_{Hl_3}\right) $\\
\hline
\rowcolor{celeste} $\delta(g^{\nu}_A)$&$\delta \bar{g}_Z \,(g^{\nu}_{A})^{SM}_{pr}  - \frac{1}{4 \, \sqrt{2} \, \hat{G}_F}
\left(C_{Hl_1} - C_{Hl_3} \right)  $\\
\hline
\rowcolor{verdino} $\delta (g^{u}_V)$&$\delta \bar{g}_Z \, (g^{u}_{V})^{SM}_{pr}  +
\frac{1}{4 \, \sqrt{2} \, \hat{G}_F} \left(- C_{Hq_1} + \, C_{Hq_3} -C_{Hu} \right) + \frac{2}{3} \delta s_W^2$\\
\hline
\rowcolor{rosa} $\delta(g^{u}_A)$&$\delta \bar{g}_Z \, (g^{u}_{A})^{SM}_{pr}
-\frac{1}{4 \, \sqrt{2} \, \hat{G}_F} \left(C_{Hq_1} - \, C_{Hq_3} -C_{Hu} \right)$  \\
\hline
\rowcolor{celeste} $\delta (g^{d}_V)$&$\delta \bar{g}_Z \,(g^{d}_{V})^{SM}_{pr}
-\frac{1}{4 \, \sqrt{2} \, \hat{G}_F} \left(C_{Hq_1} + \, C_{Hq_3} + C_{Hd} \right) -  \frac{1}{3} \delta s_W^2$ \\
\hline
\rowcolor{verdino} $\delta(g^{d}_A)$&$\delta \bar{g}_Z \,(g^{d}_{A})^{SM}_{pr}
+\frac{1}{4 \, \sqrt{2} \, \hat{G}_F} \left(-C_{Hq_1} - \, C_{Hq_3} + C_{Hd} \right)$ \\ 
\hline
\rowcolor{rosa} $\delta(g^{W_{\pm},\ell}_V)= \delta(g^{W_{\pm},\ell}_A)$  & $ \frac{1}{2\sqrt{2} \hat{G}_F} \left(C_{Hl_3} + \frac{1}{2} \frac{c_W}{ s_W} \, C_{HWB} \right)
+ \frac{1}{4} \frac{\delta s_W^2}{s^2_W}$ \\
\hline
\rowcolor{celeste} $\delta(g^{W_{\pm},q}_V)=\delta(g^{W_{\pm},q}_A)$ & $\frac{1}{2\sqrt{2} \hat{G}_F} \left(C_{Hq_3} + \frac{1}{2} \frac{c_W}{ s_W} \, C_{HWB} \right)
+ \frac{1}{4} \frac{\delta s_W^2}{s^2_W}$ \\
\hline
\hline
\end{tabular}
\caption{Anomalous fermion couplings at LO }
\label{tab:coupling-shifts}
\end{table}

For the charged currents, we define
\bea
\mathcal{L}_{W,eff} = - \frac{ \sqrt{2 \,\pi \, \hat{\alpha_e}}}{s_W} \left[(J_{\mu}^{W_\pm, \ell})_{ij} W_\pm^\mu + (J_{\mu}^{W_\pm, q})_{ij} W_\pm^\mu\right],
\eea
where in the SM one has
\bea
(J_\mu^{W_{+}, \ell})_{ij} &=&   \, \bar{\nu}_i \, \gamma^\mu \,\left(\bar{g}^{W_{+},\ell}_V - \bar{g}^{W_{+},\ell}_A \gamma_5 \right)\, \ell_j, \\
(J_\mu^{W_{-}, \ell})_{ij} &=& \, \bar{\ell}_i \, \gamma^\mu \, \left(\bar{g}^{W_{-},\ell}_V - \bar{g}^{W_{-},\ell}_A \gamma_5 \, \right) \nu_j.
\eea

The contribution of these shifts to the observables of our interest can then be calculated \cite{Falkowski:2014tna,Falkowski:2015krw,Berthier:2015gja,Berthier:2015gja,Brivio:2017vri,Falkowski:2017pss,Dawson:2019clf}.
\section{$W$ and $Z$ pole observables of interest}
\label{sec:zpole}
In this section, we briefly discuss the contributions from SMEFT to the observables considered.
Our list of EWPO includes~\cite{deBlas:2022hdk, Dawson:2019clf,Anisha:2020ggj,ALEPH:2005ab}:
\begin{eqnarray}
&&M_W, \Gamma_W, \Gamma_Z,  \sigma_h,  R_l, A_l^{FB}, R_b, R_c,  A_b^{FB}, A_c^{FB},  A_b, A_c, A_l, A_l^{SLD}, BR_{W\rightarrow \nu l}\, . \nonumber \\
\end{eqnarray}


In the SMEFT, at tree level, the decay width of $Z$ boson to fermions can be given by: 
\bea
 \bar{\Gamma} \left(Z \rightarrow f \bar{f} \right) &=& \frac{ \, \sqrt{2} \, \hat{G}_F \hat{m}_Z^3 \, N_c}{3 \pi} \left( |\bar{g}^{f}_V|^2 + |\bar{g}^{f}_A|^2 \right), \\
 \bar{\Gamma} \left(Z \rightarrow {\rm Had} \right) &=& 2 \, \bar{\Gamma} \left(Z \rightarrow u \bar{u} \right)+ 3  \, \bar{\Gamma} \left(Z \rightarrow d \bar{d} \right),
 \eea
with our chosen normalization of $\bar{g}_{V}^f = T_3/2 - Q^f \, \bar{s}_\theta^2$ and $\bar{g}_A^f = T_3/2$ where $T_3 = 1/2$ for $u_i,\nu_i$ and $T_3 = -1/2$ for $d_i,\ell_i$
and $Q^f = \{-1,2/3,-1/3 \}$ for $f = \{\ell,u,d\}$. Here, $N_C$ is 3 for quarks and 1 for leptons. The modification of the decay widths in the SMEFT compared to the situation in the SM introduces corrections of the form~\cite{Falkowski:2015krw, Falkowski:2014tna,Berthier:2015gja, Berthier:2015oma,Brivio:2017vri}
\bea
\delta \Gamma_{Z \rightarrow \ell \bar{\ell}}&=& \frac{\sqrt{2} \, \hat{G}_F \hat{m}_Z^3}{6 \pi} \, \left[ -\delta g^{\ell}_A + \left(-1 + 4 s_W^2 \right) \delta g^{\ell}_V \right] 
, \\
\delta \Gamma_{Z \rightarrow \nu \bar{\nu}}&=& \frac{\sqrt{2} \, \hat{G}_F \hat{m}_Z^3}{6 \pi} \, \left[ \delta g^{\nu}_A +  \delta g^{\nu}_V \right] 
, \\
\delta \Gamma_{Z \rightarrow Had}&=& 2 \, \delta \Gamma_{Z \bar{u} u} + 3 \, \delta \Gamma_{Z \bar{d} d}, \\ &=& \frac{ \, \sqrt{2} \, \hat{G}_F \hat{m}_Z^3}{\pi} \left[ \delta g^{u}_A - \frac{1}{3} \left(- 3 + 8 s_W^2 \right) \delta g^{u}_V - \frac{3}{2} \delta g ^{d}_A + \frac{1}{2}\left(- 3 + 4 s_W^2 \right) \delta g^{d}_V  \right], \nonumber \\ 
\eea
\bea
\delta \Gamma_{Z} &=& 3\delta \Gamma_{Z \rightarrow \ell \bar{\ell}} + 3 \delta \Gamma_{Z \rightarrow \nu \bar{\nu}} +\delta \Gamma_{had}, \\&=&  \frac{ \, \sqrt{2} \, \hat{G}_F \hat{m}_Z^3}{  2\, \pi} \left[ \delta g^{\nu}_A +  \delta g^{\nu}_V -\delta g^{\ell}_A + \left(-1 + 4 s_W^2 \right) \delta g^{\ell}_V, \right. \nonumber \\ &\, & \hspace{2.2cm} \left. + 2 \delta g^{u}_A - \frac{2}{3} \left(- 3 + 8 s_W^2 \right) \delta g^{u}_V - 3 \delta g ^{d}_A + \left(- 3 + 4 s_W^2 \right) \delta g^{d}_V \right], \nonumber \\ 
.
\eea
Here: $ \bar{\Gamma} \left(Z \rightarrow f \bar{f} \right)= \Gamma_{Z \rightarrow f \bar{f}}^{SM} + \delta \Gamma_{Z \rightarrow f \bar{f}}$ for all $f$ and the same kind of relation holds for  $\bar{\Gamma}_{Z \rightarrow Had}$.

In terms of the partial widths in the SM {\it at}  the $Z$ pole,
 $\sigma_{e^+e^- \rightarrow had}$ has the expression
\bea
\overline \sigma_h^0 =  12 \pi \, \frac{\overline \Gamma_{Z \rightarrow e \bar{e}} \overline \Gamma_{Z \rightarrow Had}}{|\overline \omega(M_Z^2)|^2},
\eea
 with $\overline \Gamma_{Z \rightarrow e \bar{e}}$, $\overline \Gamma_{Z \rightarrow Had}$ being the decay widths in SM. In the narrow-width approximation, $\overline \omega(M_Z^2) = \bar{M}_Z \, \bar{\Gamma}_Z$. The expression for its shift is relegated to Appendix~\ref{sec:appendix}.  

The shift of the ratios of decay rates defined in the SM as $R^0_{f}=\frac{\Gamma_{had}}{\Gamma_{Z \rightarrow \bar{f} f}}$ (where $f$ can be a charged lepton $\ell$ or a neutrino)
follows from
\bea
\delta R^0_{f}=\frac{1}{(\Gamma(Z \rightarrow f \bar{f})^2)_{SM}} \left[ \delta \Gamma_{Z \rightarrow Had} (\Gamma(Z \rightarrow f \bar{f}))_{SM} - \delta \Gamma_{Z \rightarrow f \bar{f}}  (\Gamma \left(Z \rightarrow {\rm Had})_{SM} \right)\right],
\eea

and we can then write $\bar{R}^0_{f}= R^0_{f} + \delta R^0_{f}$. For an identified quark, the inverse ratio is used.
The forward backward asymmetry for a $2\rightarrow 2$ scattering process is defined as
\bea
A_{FB} = \frac{\sigma_F - \sigma_B}{\sigma_F + \sigma_B}.
\eea
If $\theta$ is the angle between incoming lepton $\ell$ and outgoing fermion $f$ then, 
$\sigma_F$ is the cross-section defined in the region $\theta \in \left[0,\pi/2\right]$ and $\sigma_B$ is the cross-section defined in the region $\theta \in \left[\pi/2, \pi \right]$. In terms of the couplings, it can also be expressed as $A_{FB}^{0, f}= \frac{3}{4} A_{e} A_{f}$ where
\bea
\quad A_{e}= 2 \frac{g^{\ell}_V g^{\ell}_A}{ (g^{\ell}_V)^2 + (g^{\ell}_A)^2}, \quad A_{f}= 2 \frac{g^{f}_V g^{f}_A}{ (g^{f}_V)^2 + (g^{f}_A)^2}.
\eea
Once we include the SMEFT operators, the $Z$ couplings receive corrections (given in Table~\ref{tab:coupling-shifts}) which in turn, induces corrections to these observables. The expressions for the corrections pertaining to some of the observables have also been shown in Appendix~\ref{sec:appendix} in detail.

The partial $W$ width within the SMEFT is given by~\cite{Berthier:2016tkq,Brivio:2017vri}:
\begin{align}
\bar{\Gamma}_{W \rightarrow \bar f_i f_j} =& \Gamma_{W \rightarrow \bar f_i f_j}^{SM} + \delta \Gamma_{W \rightarrow \bar f_i f_j},\\
\Gamma_{W \rightarrow \bar f_i f_j}^{SM} =& \frac{N_C \, |V^f_{ij}|^2 \sqrt{2} \hat{G}_F \, \hat{m}_W^3}{12 \pi},\\
\delta\Gamma_{W \rightarrow \bar f_i f_j} =& \frac{N_C \, |V^f_{ij}|^2 \sqrt{2} \hat{G}_F \, \hat{m}_W^3}{12 \pi} \left( 4 \delta g_{V/A}^{W_\pm, f} + \frac{1}{2} \frac{\delta m_W^2}{\hat{m}_W^2}\right).
\end{align}
$V^f_{ij}$ are CKM or PMNS mixing matrix elememts as the case may be. Summing over all the modes we can compute the contribution to the total decay width.
\subsection{LEP-II data}
One issue with using just the data from EWPO is that they can only constrain 8 linear combination of the WCs out of a total of 10 WCs which affect the EWPO at tree-level, resulting in two blind directions. The origin of this can be traced to the fact that there is a reparametrisation invariance where the fields and the couplings have the property: $\mathcal{F}, g \rightarrow \mathcal{F}'(1+\epsilon)g'(1-\epsilon)$ \cite{Han:2004az,Berthier:2016tkq,Brivio:2017bnu} for the $2 \rightarrow 2$ scattering processes. In order to break this invariance, one has to go to $2 \rightarrow 4$ scattering processes. These were suppressed at LEP-I (off-shell processes) but realised in LEP-II due to the possibility of having resonant production of $W^+$ and $W^-$. Pair produced $W$'s can then further decay leptonically to ($\ell, \nu_\ell$) or hadronically to ($j, j$). In addition to breaking the invariance, measurement of these four fermions final states also provide additional observables for electroweak precision studies. The contributions from the SMEFT to these processes have been calculated for assuming both resonant and non-resonant precesses. Details of these observables and their $\chi^2$ analysis have also been studied in Ref.~\cite{Ellis:2018gqa}.

\section{Going Beyond EWPO: $\Delta_{CKM}$}
\label{sec:LEFT}
The aim of this section is to understand the SMEFT operators that enter the LEFT observables like muon decay and $\beta-$decay which are relevant in electroweak precision study. In~\cite{Cirigliano:2009wk}, the authors studied the maximal deviation of $\Delta_{CKM}$ allowed by electroweak precision measurements. Their study was specific to the combination of $\mathcal{O}_{ll}, \mathcal{O}_{lq_3}, \mathcal{O}_{Hl_3}$ and $\mathcal{O}_{Hq_3}$ SMEFT operators that contribute at tree-level to the $\Delta_{CKM}$ constraint, in the limit of $U(3)^5$ invariance. 

Since the SMEFT operators $\mathcal{O}_{HD}$ and $\mathcal{O}_{HWB}$, among others, given in Table~\ref{tab:opdef}, are crucial to electroweak study, here we ask the question regarding how much they influence $\Delta_{CKM}$ at 1-loop order. Moreover, the LHC direct bounds on these operators are not severe as those on the leptonic ones. In such a scenario, it becomes important to understand these operators and their contributions in low-energy observables. Hence, in this section, we match the muon decay and $\beta-$decay to 1-loop order~\cite{Dekens:2019ept,Jenkins:2017jig, Jenkins:2017dyc}, thus including the SMEFT operators given in Table~\ref{tab:opdef} in the $\Delta_{CKM}$ constraint. In particlular, the operator [$\mathcal{O}_{lq_3} \equiv (\bar{l}\tau^a \gamma_\mu l)(\bar{q}\tau^a \gamma^\mu q)$] apperaing at tree-level in $\Delta_{CKM}$ is neglected as it doesn't impact the EWPO if we neglect the off-shell processes.
 
Assuming a $U(3)^5$ flavour symmetry~\cite{Cirigliano:2009wk}, the correction to CKM unitarity is given by 
\begin{equation}
\Delta_{CKM} = |V_{ud}|^2 + |V_{us}|^2 + |V_{ub}|^2-1 \ .
\end{equation}
Here, $V_{ub}$ is very small and can be neglected. The CKM matrix element $|V_{ud}|$ is precisely measured from the study of super allowed $0^+ \to 0^+$ beta decay. A crucial input for the correct determination is the precision of $G_F$. To understand the New Physics contribution to $G_F$, let us begin by studying the muon decay process. 
\subsection{Muon Decay}
At low energies, the Lagrangian density that leads to the process $\mu \to e \bar{\nu}_e \nu_\mu$ is given by 
\begin{equation}
\mathcal{L}_{\mu} = L^{V,LL}_{\nu e} (\bar{\nu}_{L\mu} \gamma^\mu \nu_{L e})(\bar{e}_{L}\gamma_\mu \mu_L) + L^{V,LR}_{\nu e} (\bar{\nu}_{L\mu} \gamma^\mu \nu_{L e})(\bar{e}_R \gamma_\mu \mu_R) \ ,
\label{eq:muonlag}
\end{equation}
where, all the WCs are measured at the muon mass scale.

At tree-level, the matching of the LEFT Lagrangian with the SMEFT one gives,
\begin{eqnarray}
L^{V,LL}_{\nu e \ tree} &=& \dis -\frac{2}{v^2} + \Big(C_{ll (1,2,2,1)}+C_{ll (2,1,1,2)}) -2 (C_{Hl_3 (2,2)} + C_{Hl_3 (1,1)})\Big) U_{(1,1)}U^{\dagger}_{(2,2)},  \nonumber \\
\end{eqnarray}
where $U$ is the PMNS matrix. At 1-loop level, the matching of LEFT Lagrangian with the SMEFT Lagrangian gives,
\begin{eqnarray}
L^{V,LL}_{\nu e \ 1-loop} &=& \dis -1.48946 \log  \mu_W^2 -6.8206 \log \mu_W^2 \Big(0.024 (C_{Hl_3 (1,1)}+ C_{Hl_3 (2,2)}) \nonumber \\
&+& \dis 0.252418 C_{Hq_3 (1,1)}-0.0029 (C_{ll (1,1,2,2)}+C_{ll (2,2,1,1)}) \nonumber \\
&+& \dis 0.001 (C_{ll (1,2,2,1)}+ C_{ll (2,1,1,2)})+0.463 C_{HD}+0.0449 C_{HWB} \bar{g} \Big) \nonumber \\
&-& \dis 0.0025 (C_{Hl_1 (1,1)}+ C_{Hl_1 (2,2)})+0.0473 (C_{Hl_3 (1,1)}+C_{Hl_3 (2,2)})\nonumber \\
&+& \dis 1.0119 C_{Hq_3 (1,1)} - 0.0206 (C_{ll (1,1,2,2)}+ C_{ll (2,2,1,1)}) \nonumber \\
&+& \dis 0.0067 (C_{ll (1,2,2,1)}+C_{ll (2,1,1,2)}) + 2.0729 C_{HD}+0.3125 C_{HWB} \bar{g},\nonumber \\
\end{eqnarray}
where $\mu_W$ is the scale of the matching which has been set to 2 GeV in our analysis. Note that the Standard Model has only left handed currents, and hence $L^{V,LR}$ is generated by new physics beyond the cut-off scale $\Lambda$ and is, at best $O(\frac{v^2}{\Lambda^2})$. Thus to $O(\frac{v^2}{\Lambda^2}),$ the $(L^{V,LR})^2$ term in the differential width may be neglected. On the other hand, for the two terms in Eq.~\ref{eq:muonlag} electron helicities are opposite. This implies that the interference between the two amplitudes would suffer an additional suppression by a factor $(m_e/m_\mu)$ or smaller. 
Even then, if we look at the tree and 1-loop matching of $L^{V,LR}_{\nu e}$ operator with  the SMEFT Lagrangian, it consists of only $C_{le (2,1,1,2)}$ operator, which does not appear in EWPO and hence cannot be constrained. Thus we assume that this SMEFT operator vanishes and neglect the corresponding LEFT operator in further analysis. 

This means that,
\begin{equation}
L^{V,LL}_{\nu e} = L^{V,LL}_{\nu e \ tree} + L^{V,LL}_{\nu e \ 1-loop} = -\frac{4 G_F}{\sqrt{2}} \ ,
\label{eq:muondecayGF}
\end{equation}

\subsection{Beta Decay}

The low-energy effective Lagrangian for the process $d \to u e^- \bar{\nu}_e$ is given by,
\begin{eqnarray}
\mathcal{L}_n &=& \dis L^{V,LL}_{\nu e d u} (\bar{\nu}_L \gamma^\mu e_L)(\bar{u}_L \gamma_\mu d_L) + L^{V,LR}_{\nu e d u} (\bar{\nu}_L \gamma^\mu e_L)(\bar{u}_R \gamma_\mu d_R) + L^{S,RR}_{\nu e d u} (\bar{\nu}_L  e_R)(\bar{u}_L d_R) \nonumber \\
&+& \dis L^{S,RL}_{\nu e d u} (\bar{\nu}_L e_R)(\bar{u}_L \gamma_\mu d_R) + L^{T,RR}_{\nu e d u} (\bar{\nu}_L \sigma^{\mu \nu} e_R)(\bar{u}_L \sigma^{\mu \nu} d_R)
\end{eqnarray}
where $L^{V,LL}_{\nu e d u}, L^{V,LR}_{\nu e d u}, L^{S,RR}_{\nu e d u}, L^{S,RL}_{\nu e d u}, L^{T,RR}_{\nu e d u}$ are the LEFT WCs. 

At tree level, the matching of the above Lagrangian with the SMEFT one gives,
\begin{eqnarray}
L^{V,LL}_{\nu e d u \  tree} &=& \dis \Big(-\frac{2}{v^2}   -2 (C_{Hl_3 (1,1)} +  C_{Hq_3 (1,1)} ) \Big)U^{\dagger}_{(1,1)} V^{\dagger}_{(1,1)}
\end{eqnarray}
where $V$ is the CKM matrix. Matching this Lagrangian with the SMEFT Lagrangian, at 1-loop, level we get,
\begin{eqnarray}
L^{V,LL}_{\nu e d u \ 1-loop} &=& \dis U^{\dagger}_{11} V^{\dagger}_{11} \Big((0.0241 \log \mu_W^2 + 0.0595)C_{Hl_3 (1,1)} \nonumber \\
&-& \dis (0.0139 \log \mu_W^2\ +0.0712) C_{Hq_1(1,1)}+ (0.1007 \log \mu_W^2 \nonumber \\
&+& \dis 0.3931) C_{Hq_3 (1,1)} + (0.1338 \log \mu_W^2 + 0.592) C_{HD} \nonumber \\
&-& \dis (0.0115\log \mu_W^2 + 0.0715) C_{HWB} \bar{g_1} + 0.0007 C_{Hl_1(1,1)} \Big)\nonumber .\\
\end{eqnarray}

The contribution from $ L^{V,LR}_{\nu e d u}, L^{S,RR}_{\nu e d u}$ and $L^{T,RR}_{\nu e d u}$ do not contain any SMEFT coefficients that contribute to the EWPO. Then, up to 1-loop order,
\begin{eqnarray}
L^{V,LL}_{\nu e d u} = L^{V,LL}_{\nu e d u \ tree} + L^{V,LL}_{\nu e d u \ 1-loop} = \frac{4 G_F}{\sqrt{2}} V_{ud}^{eff}\ .
\end{eqnarray}
Using Eq.\ref{eq:muondecayGF}, we get 
\begin{eqnarray}
V_{ud}^{eff} = \frac{L^{V,LL}_{\nu e d u}}{L^{V,LL}_{\nu e}} &=& \dis  \frac{L^{V,LL}_{\nu e d u \ tree} + L^{V,LL}_{\nu e d u \ 1-loop}}{L^{V,LL}_{\nu e \ tree} + L^{V,LL}_{\nu e \ 1-loop}} \nonumber \\
&=& \dis V_{11} (1+\delta_{SMEFT \ tree} + \delta_{SMEFT \ 1-loop}) \ ,
\label{ckm-eqn}
\end{eqnarray}
\begin{equation}
\Delta_{CKM}=\delta_{SMEFT \ tree} + \delta_{SMEFT \ 1-loop},
\end{equation}
where
\begin{eqnarray}
\delta_{SMEFT \ tree} &=& \dis 2 (C_{Hl_3(1,1)}+C_{Hq_3(1,1)}) \nonumber \\
&+& (-2 C_{Hl_3(2,2)}+C_{ll(1,2,2,1)}+C_{ll(2,1,1,2)}-2 C_{Hl_3(1,1)}), \nonumber \\
\delta_{SMEFT \ 1-loop} &=& \dis -0.0029 C_{Hl_1(1,1)}- \dis (0.0026\log \mu_W^2 +0.0172) C_{Hl_3(1,1)} \nonumber \\
&+& \dis (0.0139\log \mu_W^2 +0.0711)  C_{Hq_1(1,1)} \nonumber \\
&-& \dis (0.0255\log \mu_W^2 + 0.0917)  C_{Hq_3(1,1)} \nonumber \\
&+& \dis (0.004 \log \mu_W^2 +0.0252 ) C_{HD} \nonumber \\
&+& \dis ( 0.0249 \log \mu_W^2 +0.1645 )C_{HWB} \bar{g_1}\ ,
\end{eqnarray}

Assuming flavor universal scenarios and removing indices, the contributions to SMEFT can be written as:
\begin{eqnarray}
\delta_{SMEFT \ tree} &=& \dis 2 (C_{Hl_3}+C_{Hq_3}) \nonumber \\
&+& (-2 C_{Hl_3}+C_{ll}+C_{ll}-2 C_{Hl_3}) \nonumber \\
&=& (2 C_{ll}-2 C_{Hl_3}+2 C_{Hq_3})\nonumber \ ,\\
\delta_{SMEFT \ 1-loop} &=& \dis -0.0029 C_{Hl_1}- \dis (0.0026\log \mu_W^2 +0.0172) C_{Hl_3} \nonumber \\
&+& \dis (0.0139\log \mu_W^2 +0.0711)  C_{Hq_1} \nonumber \\
&-& \dis (0.0255\log \mu_W^2 + 0.0917)  C_{Hq_3} \nonumber \\
&+& \dis (0.004 \log \mu_W^2 +0.0252 ) C_{HD} \nonumber \\
&+& \dis ( 0.0249 \log \mu_W^2 +0.1645 )C_{HWB} \bar{g_1}\ ,
\end{eqnarray}

Note that $V_{ud}^{eff}  =  V_{11}$ in the limit where all the SMEFT coefficients vanish and the tree-level result matches with~\cite{Cirigliano:2009wk}.

\section{Fitting procedure}
\label{sec:ewf}
The results of the fit involving the EWPO have a strong dependence on the input scheme\cite{Brivio:2017vri}. In our analyses, we choose the $(G_F,M_Z \text{ and } \alpha)$ scheme and the input parameters we used are as follows: 
 \begin{eqnarray}
G_F&=&1.1663787(6)\times 10^{-5} \gev^{-2}\nonumber \\
M_Z&=&91.1876\pm .0021\gev\nonumber \\
{1\over \alpha_e} &=& {137.035999139(31)} \nonumber\\
\label{eq:inputs}
\end{eqnarray}
Let the contribution of the higher dimensional SMEFT operators, along with SM, to the electroweak observables be represented as,
\begin{eqnarray}
\mathcal{O}_i^{SMEFT}&=&\mathcal{O}_i^{SM}+\delta \mathcal{O}_i^{SMEFT}~,
\label{eq:defs}
\end{eqnarray}
where $\delta \mathcal{O}_i^{SMEFT}$'s are the functions of WCs $C_i$'s. 
\begin{table}[!htb]
\begin{small}
\begin{center}
\begin{tabular}{|l|c|c|c|c|}
\hline
\rowcolor{lightgray}Measurement& Experiment &  Precise Theory & Pull \\
\hline\hline
 \rowcolor{celeste}$\Gamma_Z$(GeV) & $2.4955\pm 0.0023$ & $2.4945\pm 0.0006$& 0.42   
\\
\hline
 \rowcolor{celeste}$\sigma_h$(nb) & $41.481\pm 0.033$ &  $ 41.482\pm 0.008$ & -0.29 \\
\hline
 \rowcolor{celeste}$R_l$ & $20.767\pm 0.025$ & $ 20.749\pm 0.009$  & 0.67 \\
\hline 
 \rowcolor{celeste}$R_b$& $0.21629\pm 0.00066$ & $ 0.21582\pm 0.00002$   & 0.71\\
\hline
 \rowcolor{celeste}$R_c$ & $0.1721\pm 0.0030$ & $ 0.17221\pm 0.00003$ & -0.03 \\
\hline 
 \rowcolor{celeste}$R_{uc}$ & $0.166\pm 0.009$ & $ 0.172227\pm 0.000032$ & -0.69\\
\hline 
 \rowcolor{celeste}$A_l$ & $0.1465\pm 0.0033$ & $ 0.1468\pm 0.0003$ & -0.09\\
\hline
 \rowcolor{celeste}$A_l(SLD)$ & $0.1513\pm 0.0021$ & $ 0.1468\pm 0.0003$ & 2.12\\
\hline
 \rowcolor{celeste}$A_b$ & $0.923\pm 0.020$ & $0.92699\pm 0.00006$ & -0.19 \\ 
\hline 
 \rowcolor{celeste}$A_c$ & $0.670\pm 0.027$ & $ 0.6677\pm 0.0001$ &  0.08\\
\hline  
 \rowcolor{celeste}$A_s$ & $0.895\pm 0.020$ & $0.9356\pm 0.00004$ &  -0.44\\ 
\hline 
 \rowcolor{celeste}$A_{l,FB}$ & $0.0171\pm 0.0010$ & $ 0.01617\pm 0.00007$  &  0.92\\
 \hline
 \rowcolor{celeste}$A_{b,FB}$ & $0.0996\pm 0.0016$ & $ 0.1029\pm 0.0002$ &  -2.04\\
\hline
 \rowcolor{celeste}$A_{c,FB}$ & $0.0707\pm 0.0035$ & $ 0.0735\pm 0.0002$ & -0.79 \\
\hline
 \rowcolor{celeste}$M_W $(GeV) & $80.4133\pm 0.008$ &$ 80.360\pm 0.006$ & 5.33 \\ 
\hline
 \rowcolor{celeste}$\Gamma_W$(GeV)  & $2.085\pm 0.042$ & $ 2.0904\pm  0.0003$ &  -0.13\\
\hline
 \rowcolor{celeste}$BR_{W \rightarrow \nu l}$  & $0.1086\pm 0.0009$ & $ 0.108271\pm  0.000024$ & 0.36 \\
\hline
\hline
\hline
 \rowcolor{verdino}$\Delta_{CKM}$  & $-0.0015 \pm 0.0007$ & $0 \pm 0$ & -2.14 \\

\hline
\end{tabular}
\caption{\label{tab:expnums} Experimental and theoretical values and uncertainities of the observables. Pulls of the measurement with best theory is also provided in the last column~\cite{Workman:2022ynf,deBlas:2022hdk,Dawson:2019clf,ALEPH:2010aa}. }
\end{center}
\end{small}
\end{table}
For $O_i^{SM}$ in Eq. \ref{eq:defs}, we list, in Table \ref{tab:expnums}, the most precisely calculated values. 
The form of the $\chi^2$ is given by 
\begin{eqnarray}
\chi^2&=&\Sigma_{i,j}(\mathcal{O}_i^{exp}-\mathcal{O}_i^{SMEFT})\sigma^{-2}_{ij} (\mathcal{O}_j^{exp}-\mathcal{O}_j^{SMEFT}) \, .
\end{eqnarray} 
where the covariance matrix $\sigma^2_{ij}=\Delta_i^{exp} \rho^{exp}_{ij} \Delta_j^{exp}+\Delta_i^{th} \rho^{th}_{ij} \Delta_j^{th}$. Here $\rho^{exp}_{ij}$ is the experimental correlation matrix obtained from \cite{ALEPH:2005ab} and $\rho^{th}_{ij}$ is identity. $\Delta_i^{exp}$ and $\Delta_i^{th}$ are the experimental and theory errors of the $i^{\rm th}$ observable. Inclusion of theoretical uncertainities and the correlations among the observables lead to better constraints over the parameters of interest.

In Table  \ref{tab:expnums}, we summarize the current status of the SM theory and the experimental results, including $\Delta_{CKM}$. First we compute the $\chi^2$ including the relevant observables from Table~\ref{tab:expnums}, thereby making it a function of the WCs which we are interested in including in our fit. The $\chi^2$ function is then minimized to get the best-fit values of the  WCs. The 1$\sigma$ ranges of the WCs are calculated by marginalizing over the other parameters. This is done in the following way: for the 1$\sigma$ range of a parameter $c_i$ of our interest, we define a partial $\chi^2$ as follows.
\[\chi^2_{par}=\chi_{c_i=x_0,c_j}^2=\chi^2(x_0,c_j),\]  
where $x_0$ is a random value for the variable $c_i$. $c_j$s are the other parameters present in the $\chi^2$ over which the marginalization is performed.
We then define the following test-statistic which is a difference between the minima of the partial $\chi^2$ and the minima of the total $\chi^2$
\[min(\chi^2_{par})-min(\chi^2(c_i,c_j))\]
For each random value of $c_i$, we get some value for this new test-statistic, which can be used to obtain an interpolating function for $c_i$.
The marginalized result for the $1\sigma$ spread of $c_i$ is then obtained by demanding the value of the obtained interpolating function to be smaller than the $1\sigma$ value of the $\chi^2$ function for a single variable.

The correlation between two WCs of our interest affecting the fit can also be calculated in a similar way by constructing a partial $\chi^2$ for the two WCs and then obtaining their interpolating function through the test-statistic.

\subsection{$S$, $T$, $V$ fitting}
The effects of new heavy particles(with mass $M_{new}$) on the gauge boson self-energies can be described by three  parameters $S$, $T$, and $U$. The $U$ parameter is not very sensitive to heavy New Physics because of the presence of an extra factor of $M_Z/M_{new}$ and hence can be neglected. This $S$ and $T$ parametrisation can be mapped to the SMEFT WCs $C_{HD}$ and $C_{HWB}$ and has been studied in literature~\cite{Akhundov:2013ons,Ciuchini:2013pca,Ciuchini:2016sjh,deBlas:2016ojx}. Along with them, we append the $V$ parameter which parametrises the change in $G_F$. This parameter is crucial if we have to make the quarks and leptons interact with BSM resonances. As shown in
Eq.~\ref{eq:gdef}, this shift in $G_F$ can be mapped to the combination $2 C_{Hl_3} - C_{ll}$. The expressions for the Electroweak Observables in terms of this $S$, $T$ and $V$ parametrisation are given in Table~\ref{table:stvobsexp}. For $\Delta_{CKM}$, we infer the result by considering only the tree-level matching, where the dependence on $V$ comes from the muon decay parameters sitting in the denominator of Eq~\ref{ckm-eqn}.  As can be seen from Table~\ref{table:stvobsexp}, this $V$ parameter is very crucial for $M_W$ as it has the highest sensitivity to this parameter, which can be inferred from its largish coefficient of $V$ compared to other observables. 
\begin{table}[htb!]
\begin{center}
\includegraphics[scale=0.3]{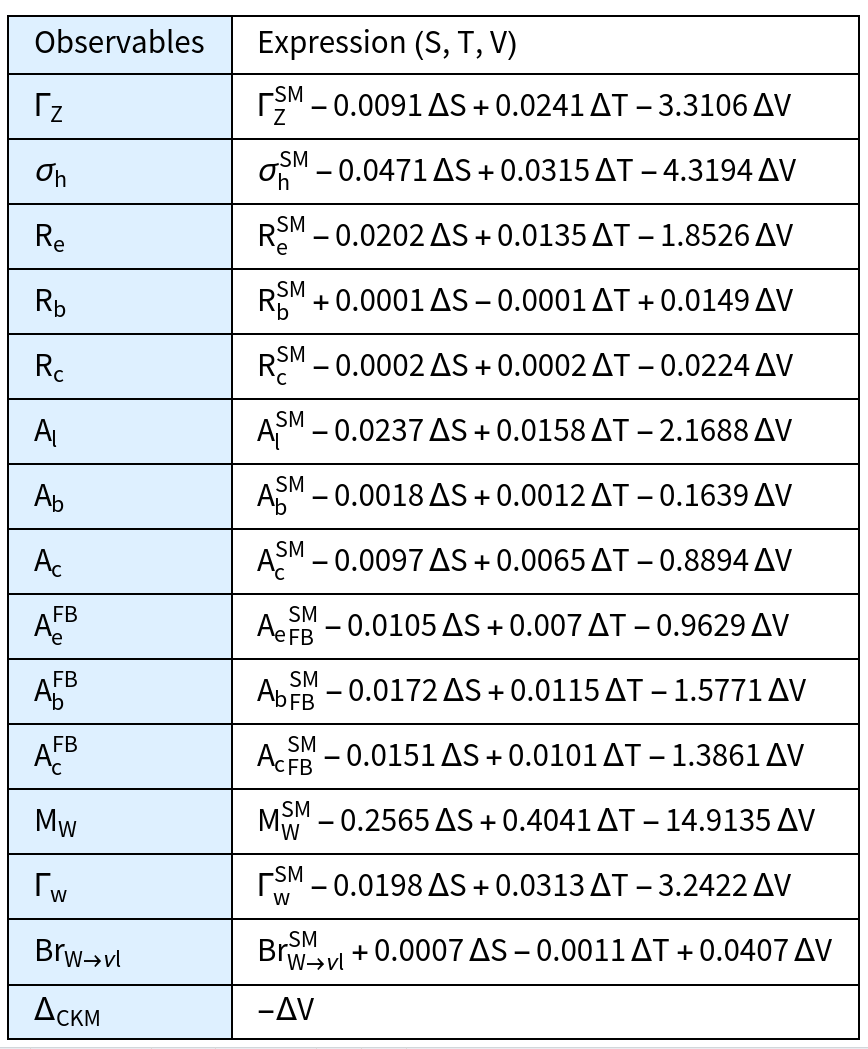}
\caption{Expressions for observables in terms of $\Delta S$, $\Delta T$ and $\Delta V$.}
\label{table:stvobsexp}
\end{center}
\end{table} 

\begin{figure}[!htb]
\begin{center}
\includegraphics[scale=0.15]{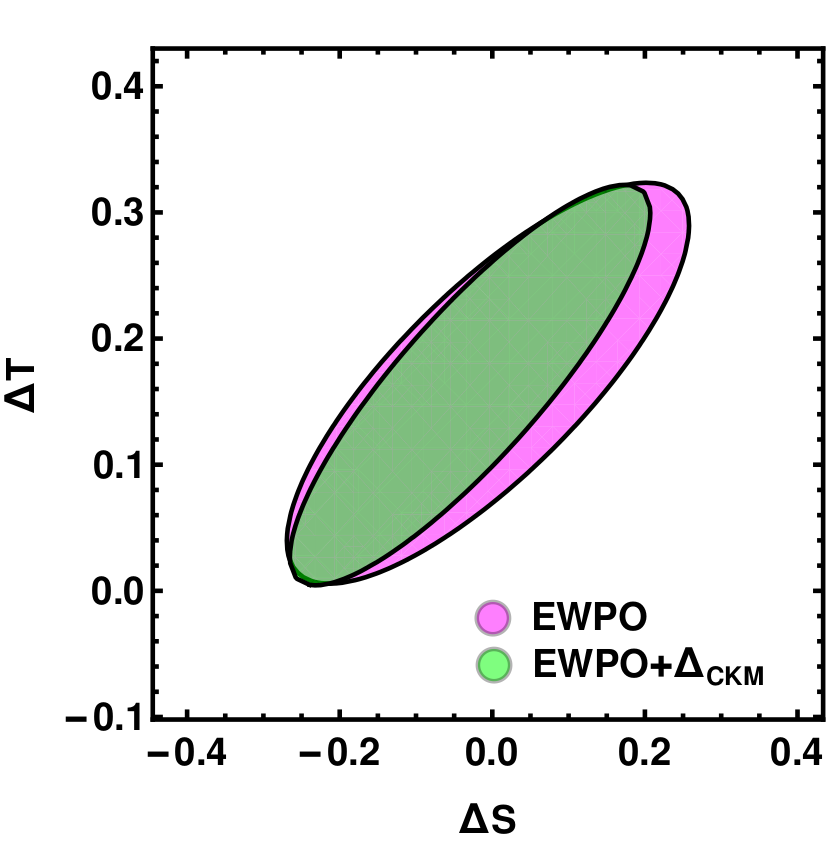}
\includegraphics[scale=0.15]{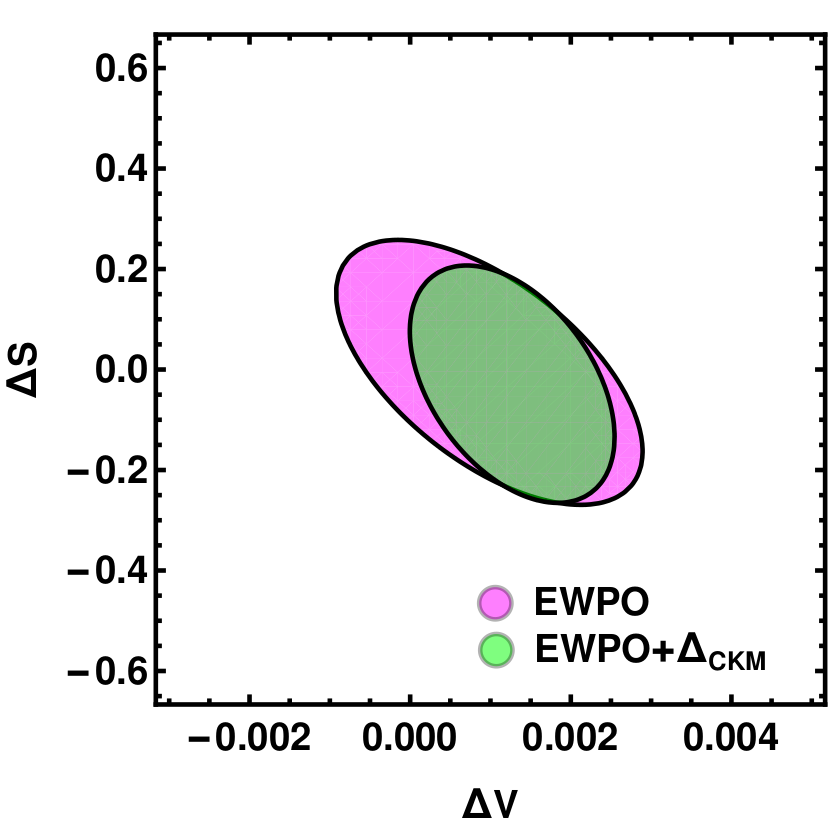}
\includegraphics[scale=0.143]{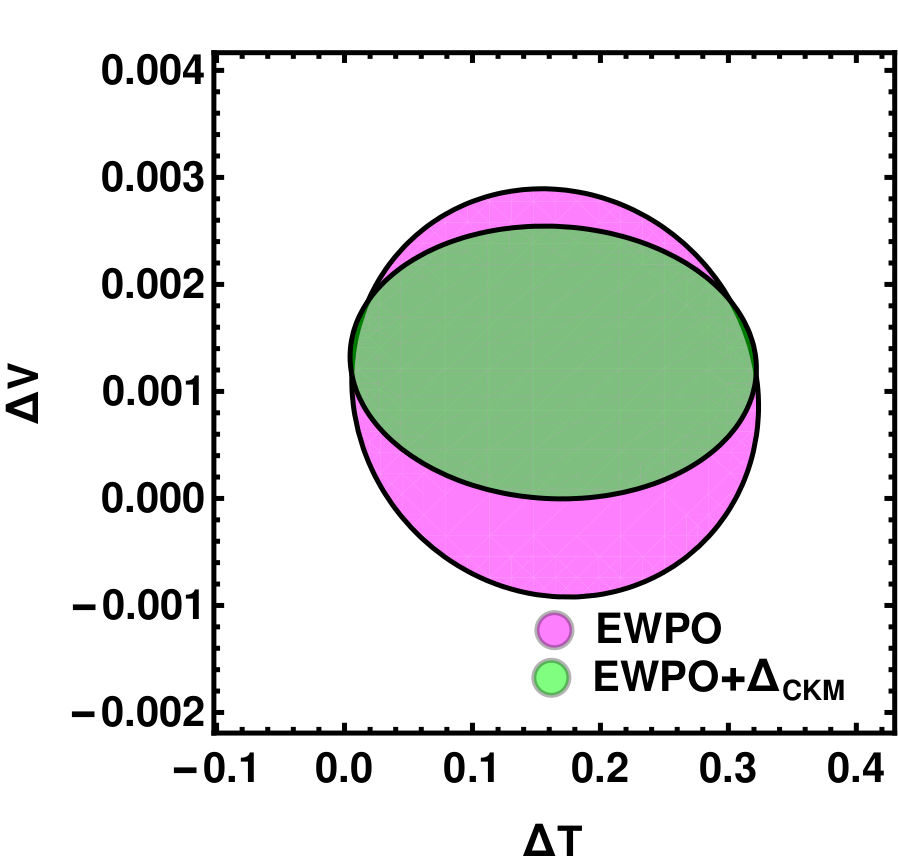}
\captionof{figure}{S, T and V parameters with EWPO and EWPO $+$ $\Delta_{CKM}$ data}
\label{fig:STV}
\end{center}
\end{figure}

In Figure~\ref{fig:STV}, we have plotted the change in $S$, $T$ and $V$ parameters with EWPO data and EWPO $+$ $\Delta_{CKM}$ data in a 2-D plane by marginalising over the third parameter. Note that the inclusion of $\Delta_{CKM}$ constrains the Peskin-Takeuchi ($S$,$T$) parameters better in comparison with just the EWPO data. The results of the fit are presented in Table~\ref{tab:STV}.
\begin{table}
\vspace{-0.2cm}
\begin{small}
    \centering
   { \begin{tabular}{|c|c|rrr|c|rrr|}
 \hline
 WC & B.F(\scriptsize{EWPO}) & \multicolumn{3}{c|}{Correlation} & B.F(\scriptsize{EWPO+CKM}) & \multicolumn{3}{c|}{Correlation} \\\hline
 & \multicolumn{4}{c|}{\small{($\chi^2_{\rm fit}$/$\chi^2_{\rm SM}=11.41/40.08$)}} & \multicolumn{4}{c|}{\small{($\chi^2_{\rm fit}$/$\chi^2_{\rm SM}=11.68/44.67$)}} \\
 \hline 
$\Delta {S}$ & $-0.0032 \pm 0.1077 $ & $1.00$ & & & $-0.0292 \pm 0.0965 $ & $1.00$ & &\\ 
$\Delta {T}$ & $ 0.1646\pm 0.0649 $ & $0.79$ & $1.00$ & &$ 0.1631\pm 0.06484 $ & $0.86$ & $1.00$ &  \\ 
$\Delta {V}$ & $-0.001\pm 0.0007$ & $-0.59$ & $-0.07$ & $1.00$&  $0.0013\pm 0.0005$ & $-0.44$ & $-0.04$ & $1.00$ \\ \hline
    \end{tabular}}
    \caption{Global fit of ($S$,$T$,$V$) paramatrisation tree-level to EWPO and $\Delta_{CKM}$.}
\label{tab:STV}
\end{small}
\end{table}
\begin{table}[!htb]
\begin{center}
\includegraphics[scale=0.29]{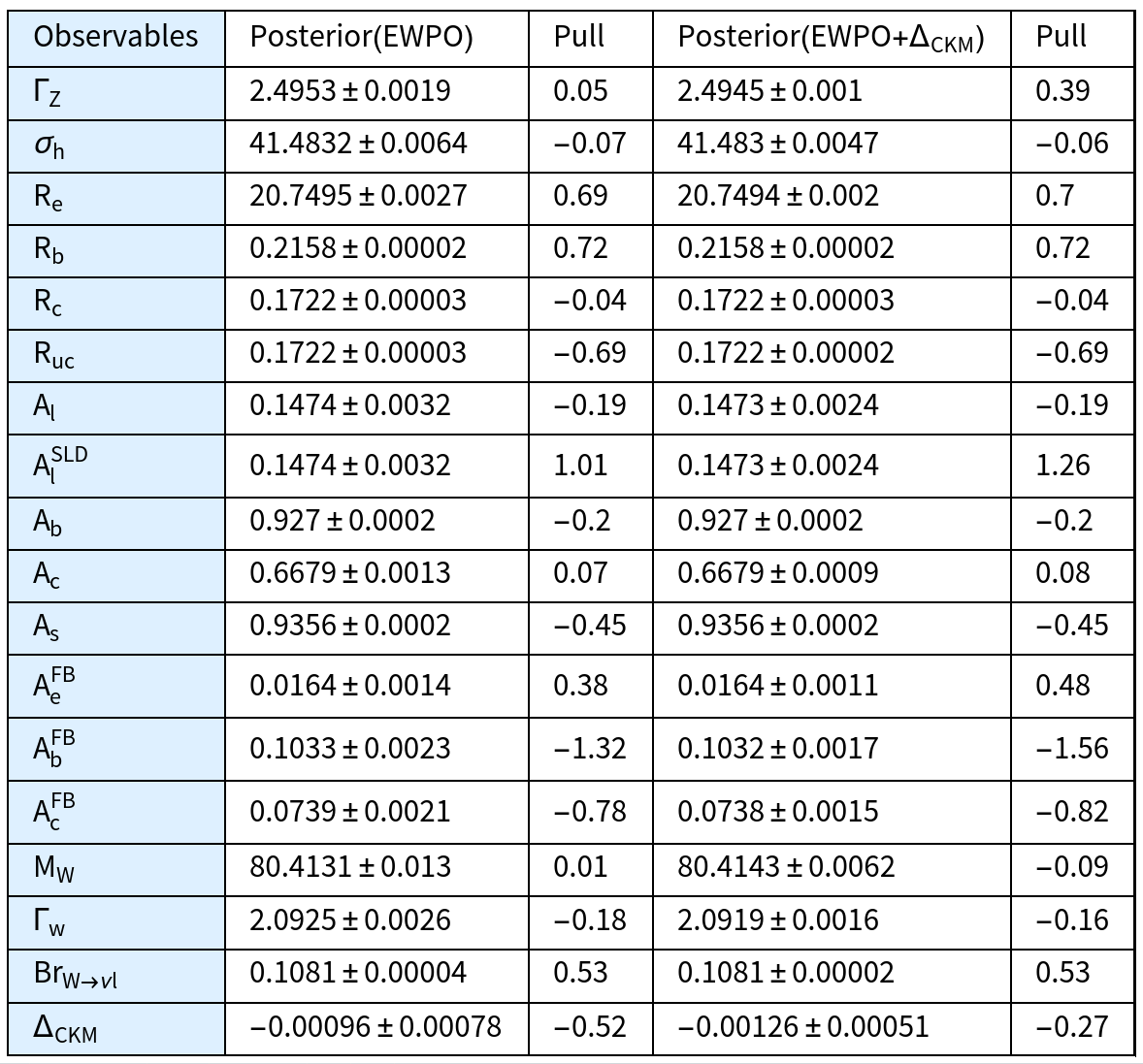}
\caption{Observables posteriors using STV fit. }
\label{table:posterior-stv}
\end{center}
\end{table}
The posterior values of the observables after the $(S,T,V)$ can be read off from  Table~\ref{table:posterior-stv}. Even without including the $\Delta_{CKM}$ constraint, we can see that this paramatrisation has very good agreement with the experimental results of $M_W$ and $\Delta_{CKM}$ simultaneously. The discrepancy in $A_l^{SLD}$ and $A_b^{FB}$ also moves below 2$\sigma$, thereby making this a paramatrisation worth considering to look for BSM physics. Inclusion of $\Delta_{CKM}$ in the fit improves its agreement with the measured value, while slightly deteriorating $A_l^{SLD}$ and $A_b^{FB}$. 

\subsection{Study of $C_{Hl_3}$ and $C_{ll}$}
As already seen from the expression of the recent anomalies in terms of SMEFT WCs, all tree-level $C_{Hl_3}$ and $C_{ll}$ are the ones which are common both in $M_W$ and $\Delta_{CKM}$ (using 1-loop matching result of $\Delta_{CKM}$). Incidentally, these WCs are also the ones which affect the weak coupling constant $G_F$. However, along with changing a $G_F$, there is an additional dependence on the WC $C_{Hl_3}$ coming from the shift in couplings for the observables as shown in Table~\ref{tab:coupling-shifts}. Similar to the $(S,T,V)$ fit, we present two distinct cases, where first we have shown the fit to the EWPO. We then augment the number of observables by including $\Delta_{CKM}$ and then present the corresponding pulls of the observables after each step. 
\begin{figure}[!htb]
\begin{center}
\includegraphics[scale=0.15]{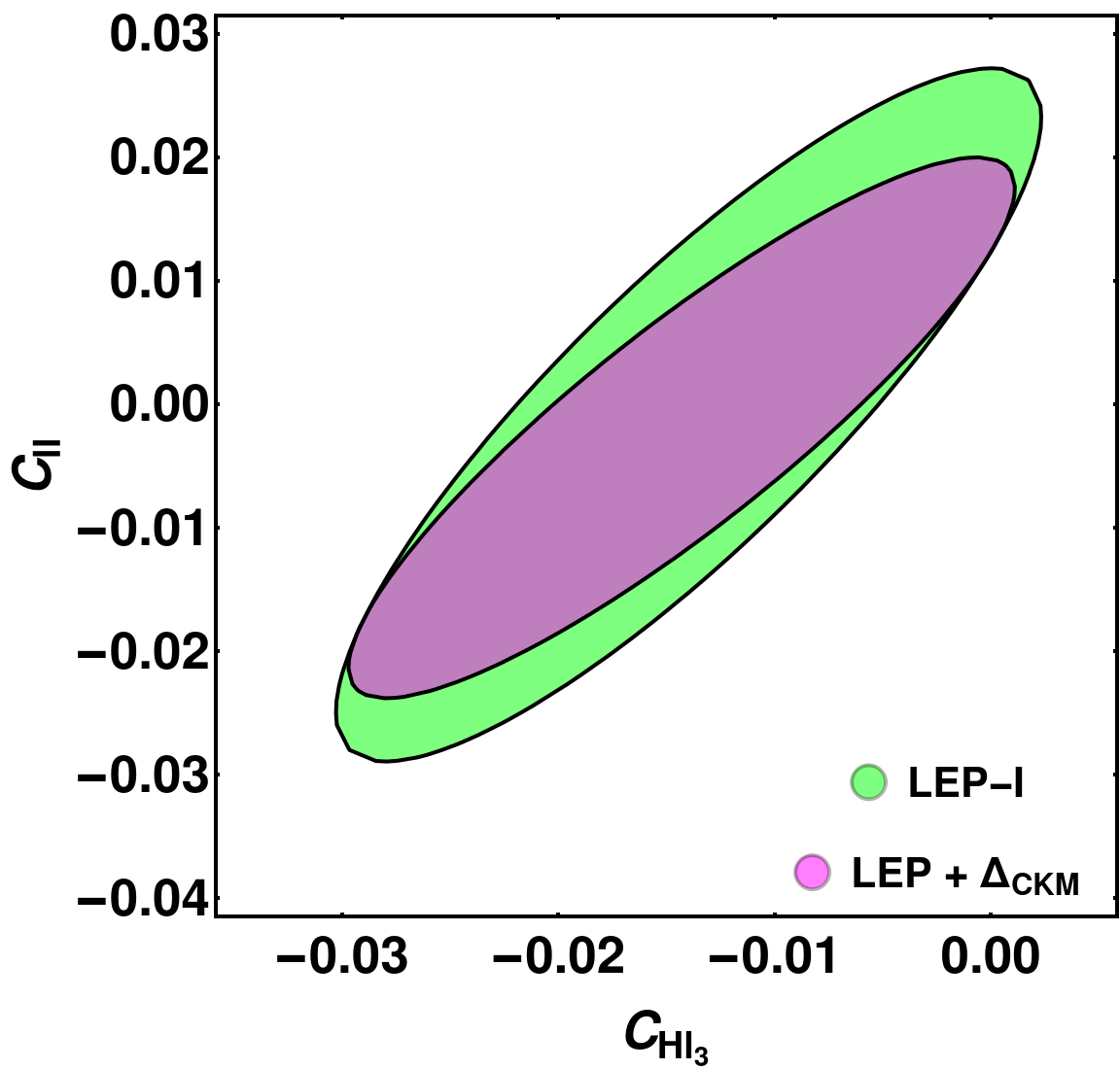}
\vspace{-0.4cm}
\end{center}
\caption{Marginalised 2-D plot for WCs common to $W$-mass and $\Delta_{CKM}$.}
\label{fig:common_WC}
\end{figure}
\begin{table}[!htb]
\begin{small}
    \centering
   { \begin{tabular}{|c|c|rr|c|rr|}
 \hline
 WC & B.F(\scriptsize{EWPO}) & \multicolumn{2}{c|}{Correlation} & B.F(\scriptsize{EWPO+CKM}) & \multicolumn{2}{c|}{Correlation} \\\hline
 & \multicolumn{3}{c|}{\scriptsize{($\chi^2_{\rm fit}$/$\chi^2_{\rm SM}=24.27/40.08$)}} & \multicolumn{3}{c|}{\scriptsize{($\chi^2_{\rm fit}$/$\chi^2_{\rm SM}=105.36/126.37$)}} \\
 \hline 
$C_{Hl_3}$ & $-0.0138\pm 0.0065$  & $1.00 $ &  & $- 0.0152 \pm 0.0058 $ & $1.00$ &  \\ 
$C_{ll}$ & $-0.0008\pm 0.0114$  & $0.86$ & $1.00$ & $-0.0034\pm 0.0084$ & $0.88$ &  $1.00$ \\ \hline
    \end{tabular}}
    \caption{Global fit of the WCs common to $W$-mass and $\Delta_{CKM}$ (using 1-loop matching result of $\Delta_{CKM}$).}
    \label{tab:common-WC}
\end{small}
\end{table}
\begin{table}[htb!]
\begin{center}
\includegraphics[scale=0.28]{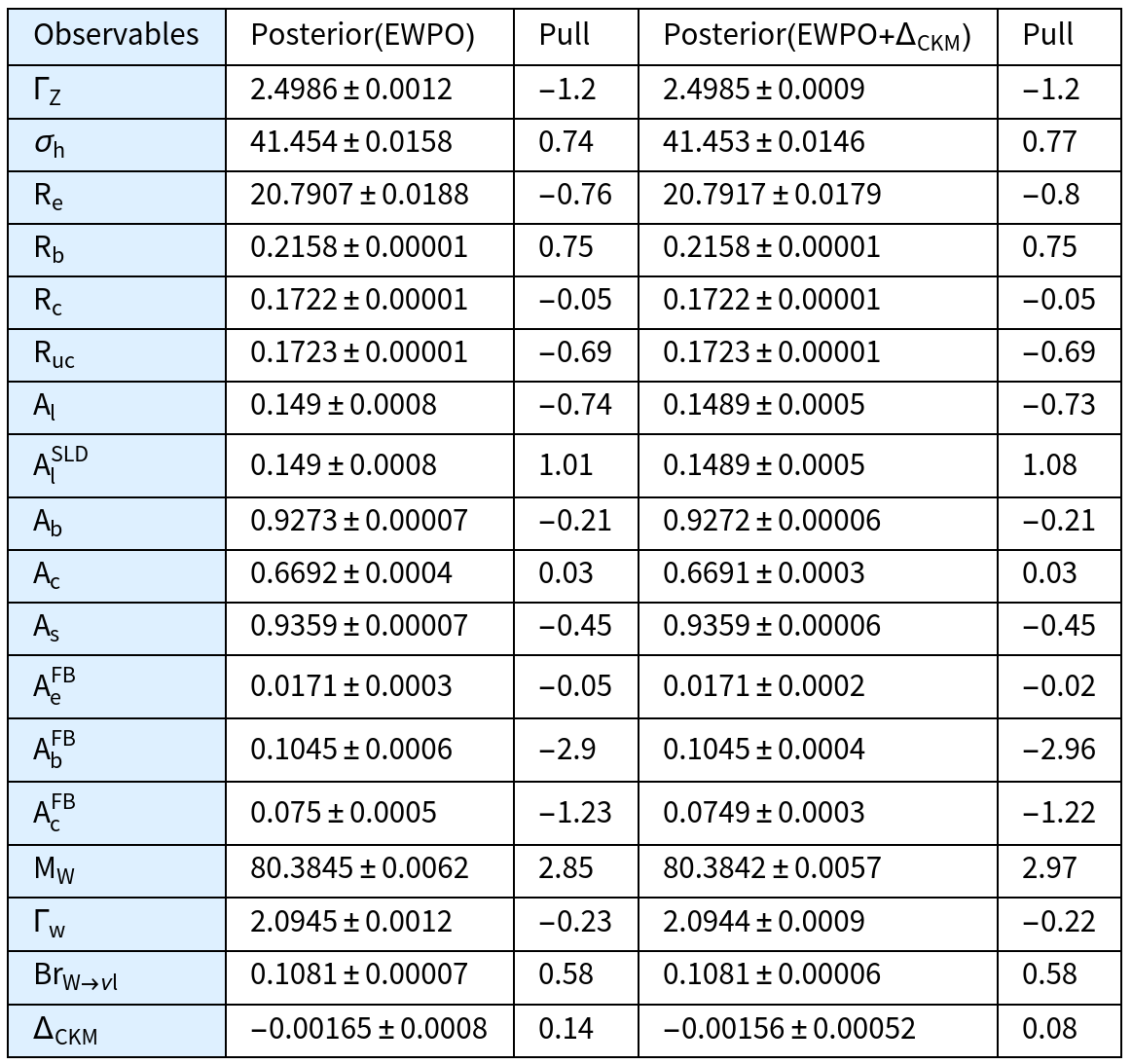}
\caption{Posterior results and Pull for WCs common to $M_W$ and $\Delta_{CKM}$, once just with EWPO observables and once including the $\Delta_{CKM}$ constraint.}
\label{table:common-WCs-post}
\end{center}
\vspace{-0.4cm}
\end{table}
As seen from Figure~\ref{fig:common_WC}, the inclusion of $\Delta_{CKM}$ has a more constraining effect on the two WCs. Table~\ref{tab:common-WC} presents the result of the fit and correlation among the WCs. The posterior results for the observables are given in Table~\ref{table:common-WCs-post}. For the first case with only the EWPO observables in the fit, the discrepancy in $M_W$ comes down to 2.85$\sigma$. The discrepancy in $A_b^{FB}$ however deteriorates to $-2.9\sigma$. Discrepancy in $\Delta_{CKM}$ however comes down to a mere 0.14$\sigma$. Inclusion of the $\Delta_{CKM}$ constraint in the fit does not significant constrain it. Thus one can argue that any New Physics effect coming through these two parameters (or $\delta G_F$) cannot fully satisfy the EWPO data. This also strengthens the case for the $(S,T,V)$ paramatrisation, where presence of extra parameters $S$ and $T$ (or equivalently $C_{HD}$ and $C_{HWB}$) help us explain all the observables within 2$\sigma$ discrepancy.

\subsection{Study of $C_{HD}$, $C_{HWB}$, $C_{Hl_3}$ and $C_{ll}$}
\begin{figure}[!htb]
\begin{center}
\includegraphics[scale=0.12]{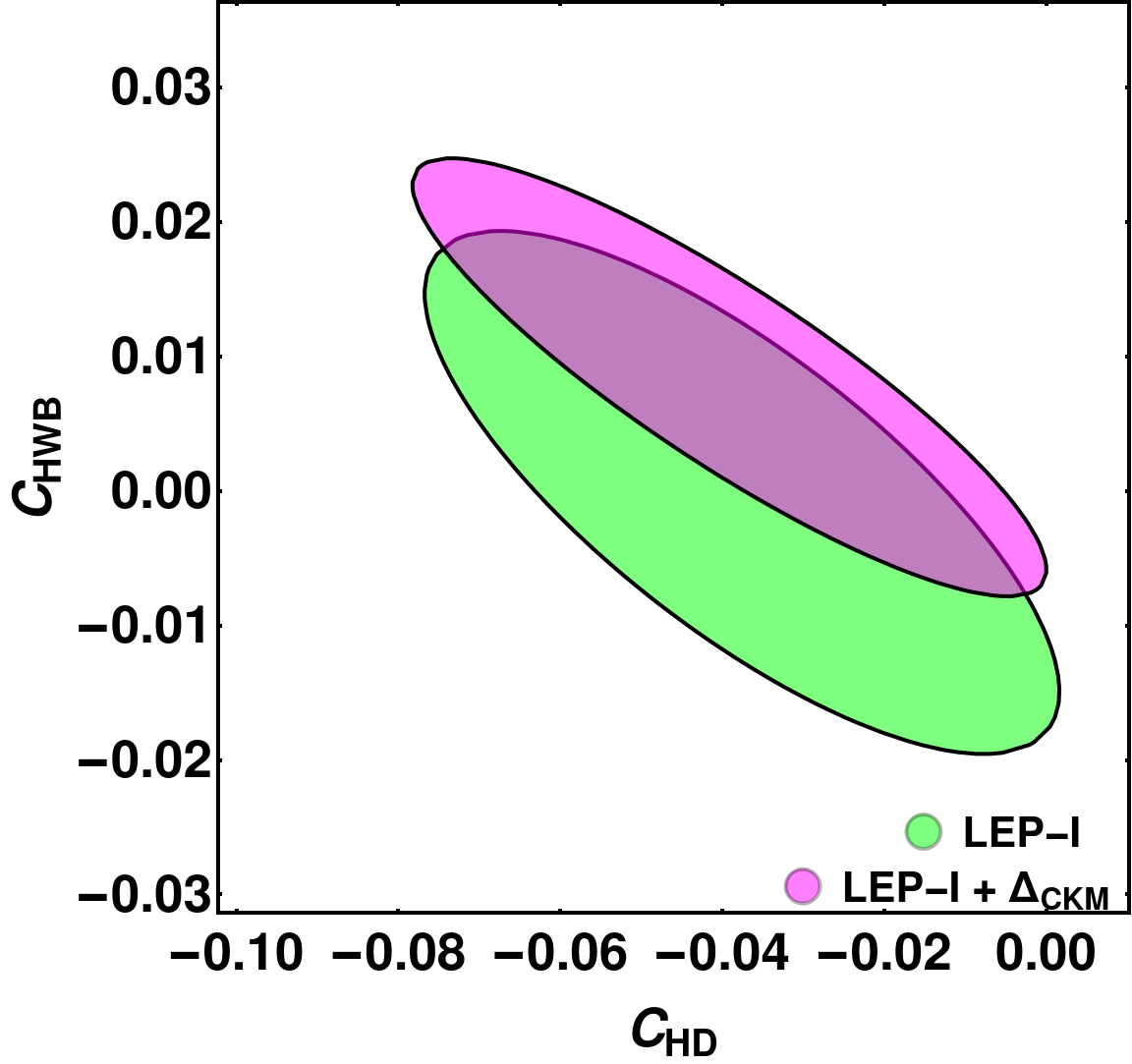}~
\includegraphics[scale=0.12]{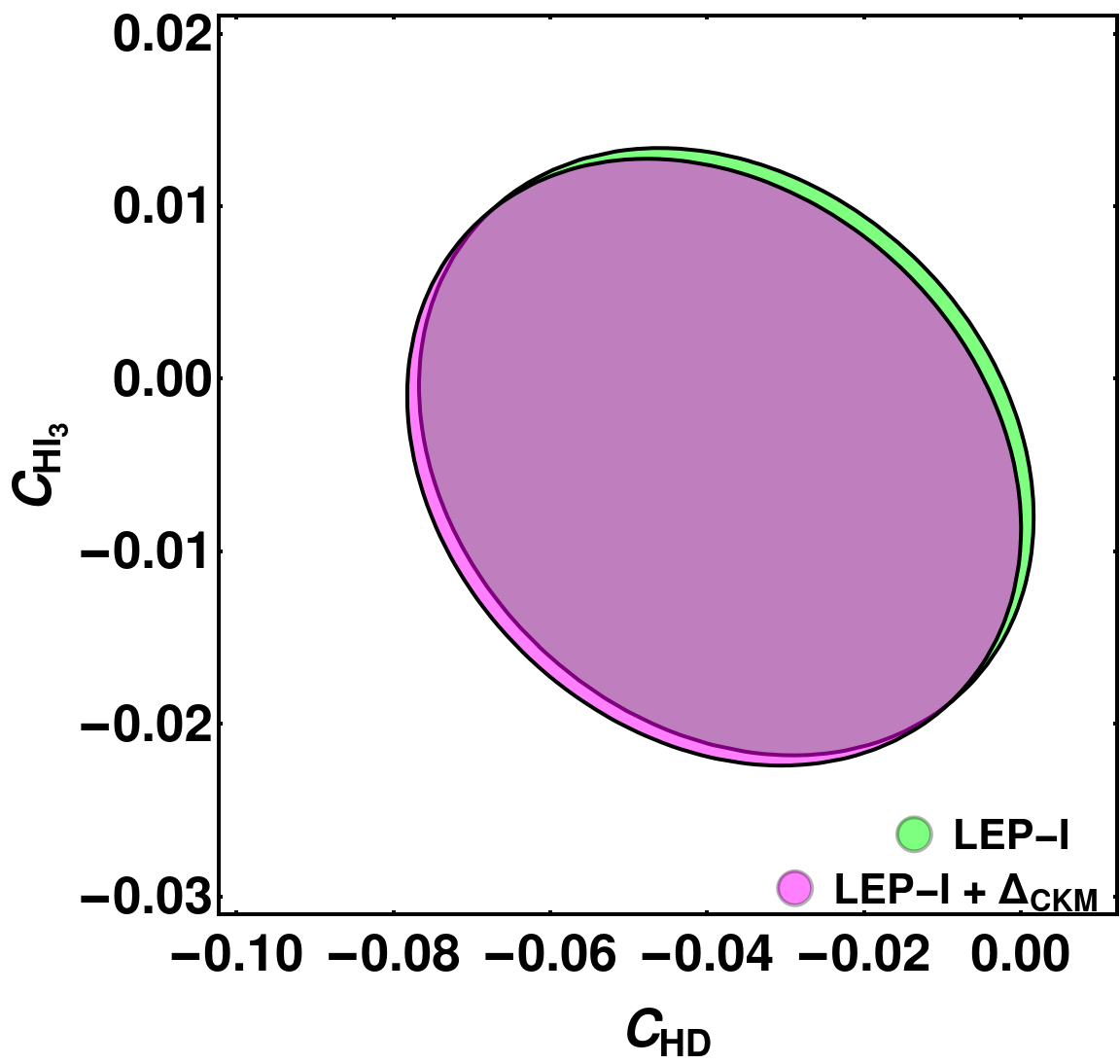}~
\includegraphics[scale=0.12]{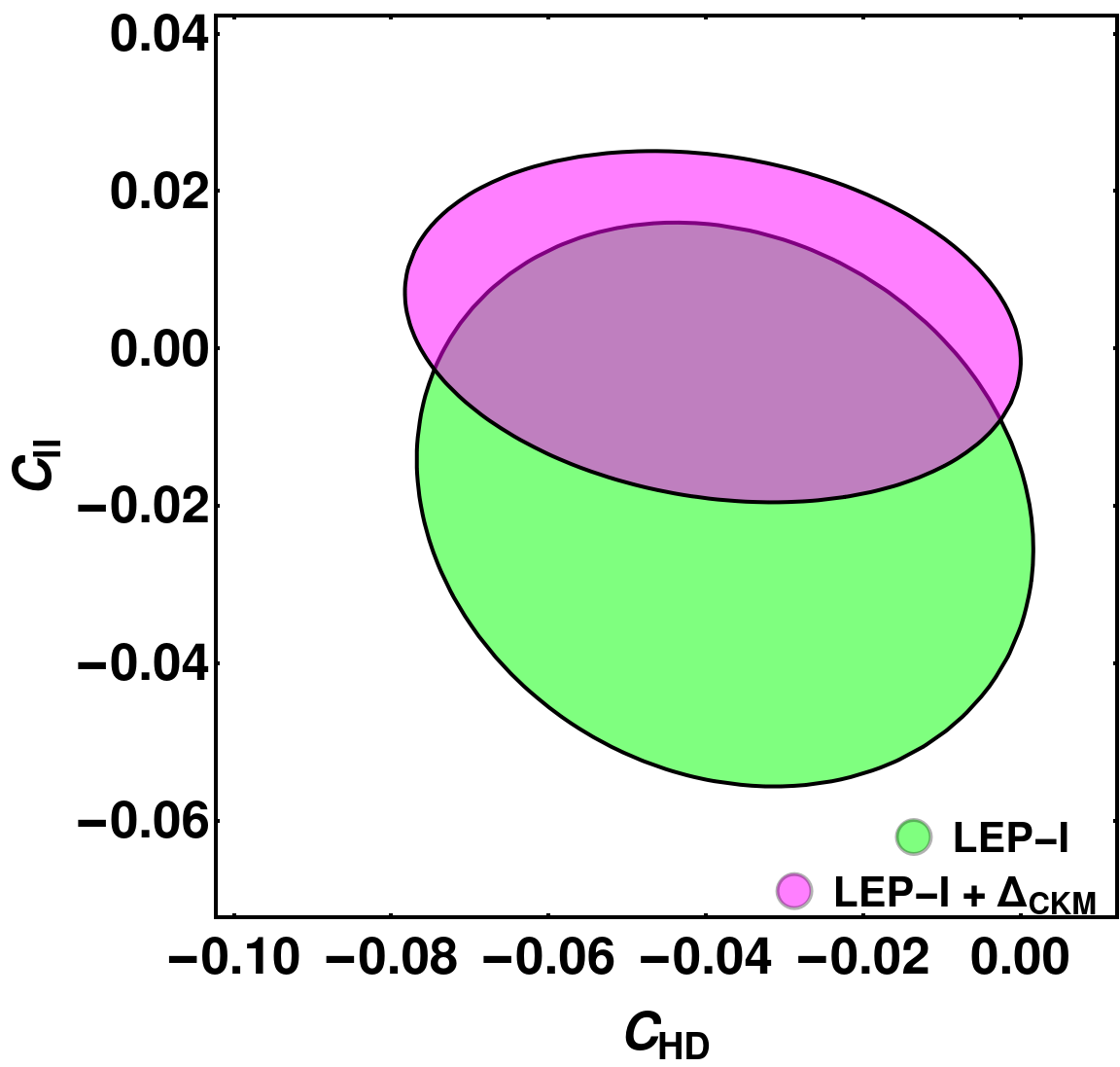}
\includegraphics[scale=0.12]{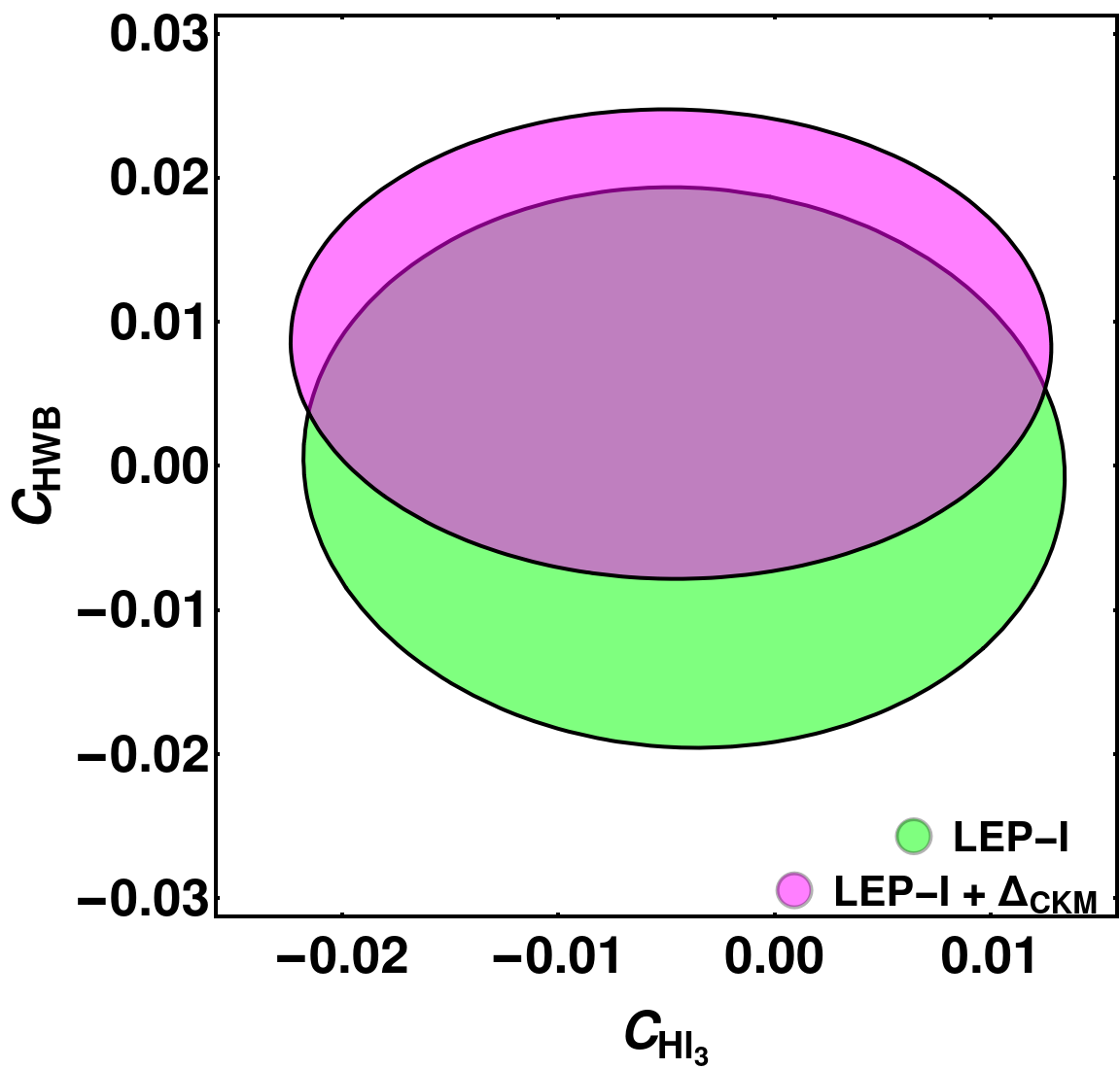}~
\includegraphics[scale=0.12]{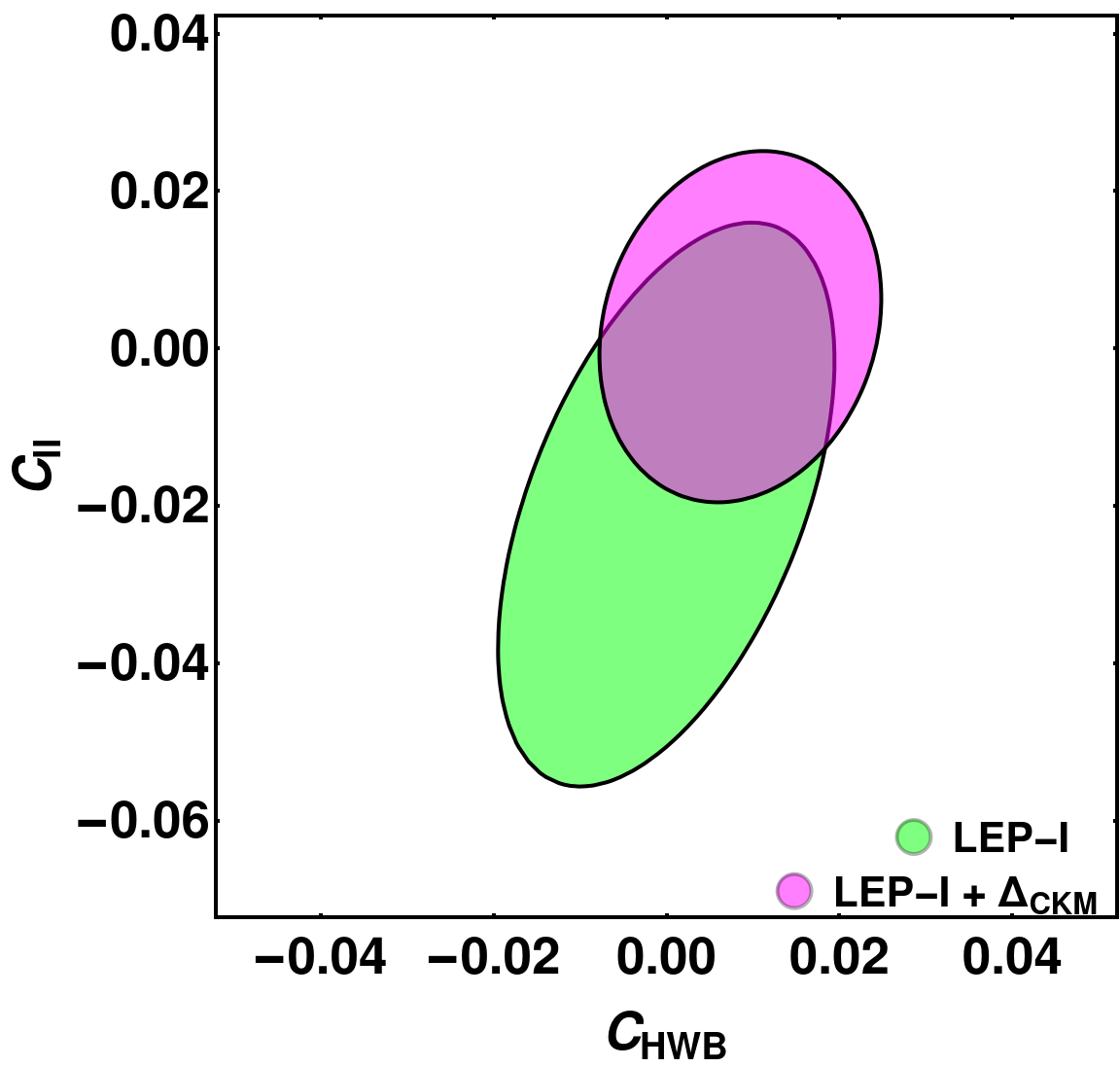}~
\includegraphics[scale=0.12]{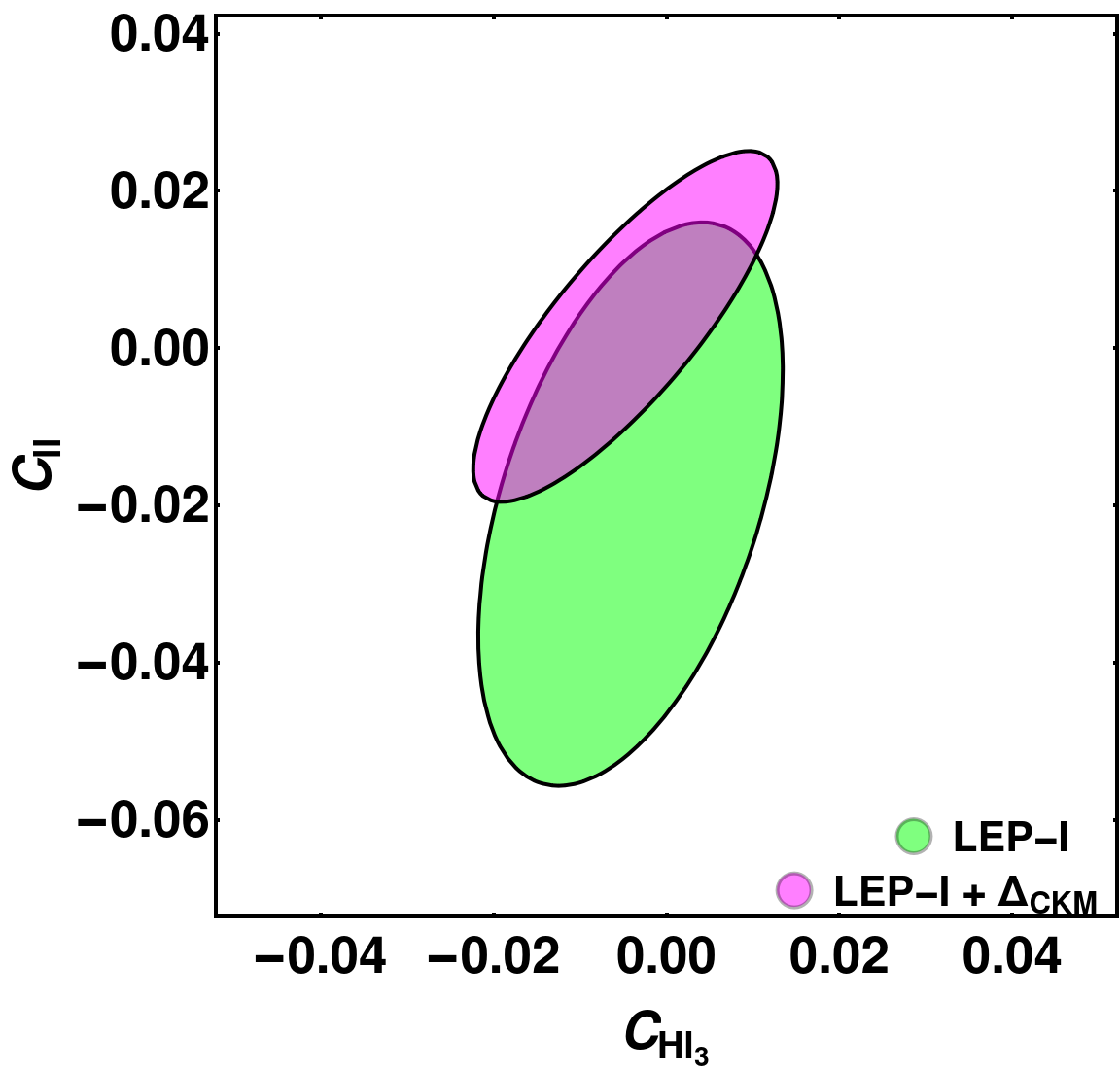}
\end{center}
\caption{2-D marginalised global fit of the four WCs affecting $M_W$ at tree level.}
\label{fig:4common_WC:W}
\end{figure}

With the result from CDF impacting the world-average of W-mass, the discrepancy now stands at around 5.3$\sigma$ from the SM value. Taking it as a sign of BSM physics, we look at a fit with just the SMEFT operators contributing to W-mass at tree-level. The shift in W mass in SMEFT can be written as:
\begin{equation}
	\frac{\delta M_W^2}{M_W^2} = 
	v^2 \ \frac{s_W c_W}{s_W^2 - c_W^2} 
	\left[ 2 \, C_{HWB} + \frac{c_W}{2 s_W} \, C_{HD} + 
	\frac{s_W}{c_W} \left( 2 \, C_{Hl}^{(3)} - C_{ll} \right) \right] \,,
\label{eq:mW}
\end{equation}
This can also be thought of as augmenting the previous fit by including two more parameters $C_{HD}$ and $C_{HWB}$. The results of the fit and the 2-D marginalised plots are shown in Table~\ref{tab:tree-W} and Figure~\ref{fig:4common_WC:W}. The plot here shows that the inclusion of the $\Delta_{CKM}$ constraint (matched to SMEFT at 1-loop level) has significant impact on the results of the fit with $C_{HWB}$ and $C_{ll}$ seeing noticeable shifts in their best fit points and their 2$\sigma$ reaches. The correlations among the WCs also show significant changes, with the $C_{ll}$-$C_{HWB}$ and $C_{ll}$-$C_{Hl_3}$ pairs showing the highest change.
\begin{table}[!htb]
\begin{scriptsize}
    \centering
 \begin{tabular}{|c|c|rrrr|c|rrrr|}
 \hline
 WC & B.F(\scriptsize{EWPO}) & \multicolumn{4}{c|}{Correlation} & B.F(\scriptsize{EWPO+CKM}) & \multicolumn{4}{c|}{Correlation} \\\hline
 & \multicolumn{5}{c|}{\scriptsize{($\chi^2_{\rm fit}$/$\chi^2_{\rm SM}=11.07/40.08$)}} & \multicolumn{5}{c|}{\scriptsize{($\chi^2_{\rm fit}$/$\chi^2_{\rm SM}=14.96/44.67$)}} \\
\hline 
$C_{HD}$ & $-0.0376 \pm 0.0159 $ & $1.00$ & & & & $-0.0392 \pm 0.0159 $ & $1.00$ & & &\\ 
$C_{HWB}$ & $ -0.00007\pm 0.0079 $ & $-0.76$ & $1.00$ & & &$ 0.0085\pm 0.0066 $ & $-0.87$ & $1.00$ & & \\ 
$C_{Hl_3}$ & $-0.0042\pm 0.0072$ & $-0.21$ & $-0.04$ & $1.00$&  & $-0.0048\pm 0.0072$ & $-0.21$ & $-0.01$ & $1.00$ &\\ 
$C_{ll}$ & $-0.0198\pm 0.0146$ & $-0.15$ & $0.51$ & $0.47$& $1.00$ & $0.0028\pm 0.0091$ & $-0.19$ & $0.16$ & $0.81$& $1.00$\\ \hline
\end{tabular}
\caption{Global fit of the WCs affecting the W-boson mass at tree-level}
\label{tab:tree-W}   
\end{scriptsize} 
\end{table}
\begin{table}[htb!]
\begin{center}
\includegraphics[scale=0.29]{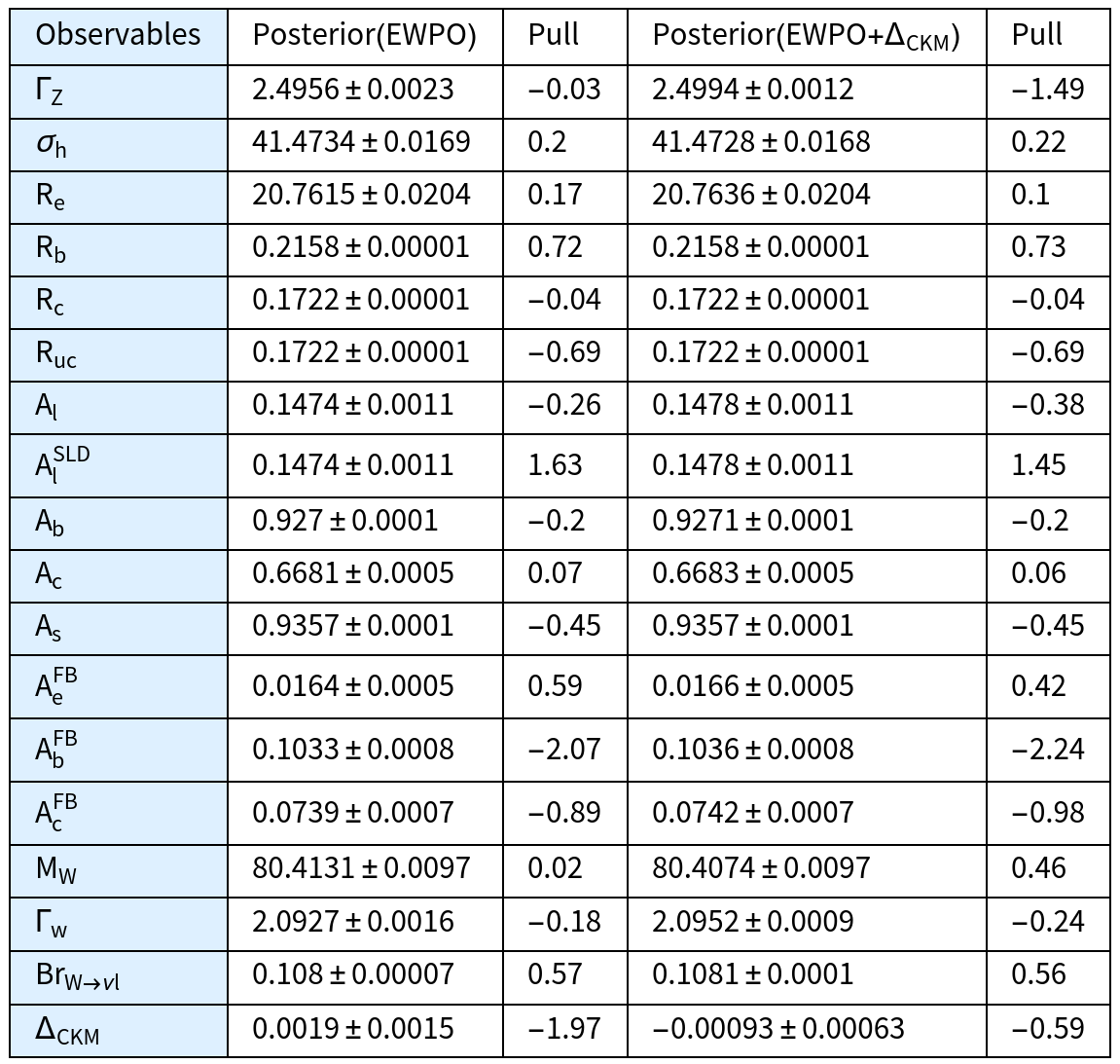}
\caption{Posterior results and Pull for WCs effecting $M_W$, once just with EWPO observables and once including the $\Delta_{CKM}$ constraint.}
\label{table:4W_params_post}
\end{center}
\end{table} 

The posterior values for the observables are given in Table~\ref{table:4W_params_post}. For the fit with EWPO, the discrepancy in $M_W$, as expected, is drastically reduced. $A_l^{SLD}$ also comes down under 2$\sigma$. However   $A_b^{FB}$ and $\Delta_{CKM}$ shows the highest discrepancy with a deviation of $-2.09\sigma$ and $-1.97\sigma$ respectively. Inclusion of the $\Delta_{CKM}$ constraint along with EWPO reduces it's discrepancy significantly to $-0.59\sigma$ at the expense of worsening $A_b^{FB}$ to $-2.24\sigma$. The agreement of $\Gamma_Z$ also suffers with its discrepancy increasing from $-0.03\sigma$ to $-1.49\sigma$. This study thus indicates that we need to go beyond these four WCs in order to satisfy the precision data.

\subsection{VLL inspired study}
Inspired by various see-saw like scenarios that admit the possibility to address other shortcomings of the SM like muon $(g-2)$, neutrino mass etc., we take a closer look at various Vectorlike leptons multiplets by identifying the leading dimension-6 operators which get affected. The list of leptons along with their corresponding Yukawa coupling $\lambda_l$ (where $l$ represents the corresponding vectorlike lepton) and the masses of the heavy states $M_l$  are shown in Table~\ref{tab:listf}. As seen from the table, the tree-level imprints of the model parameters on the SMEFT WCs are restricted to $C_{He}$, $C_{Hl_1}$  and $C_{Hl_3}$. 
\begin{table}[htb!]
\begin{small}
$$\begin{array}{|cccccc|}\hline
\rowcolor[gray]{0.8}
\hbox{VLF} & (SU(2)_L , Y)& Interaction  & C_{He} & C_{Hl_1} & C_{Hl_3}   \cr
\hline\hline
 \rowcolor{celeste}N^a  & (3,0)& N^a (H\epsilon \tau^a L) & 0 &+\lambda_N^2/4M_N^2 & -\lambda_N^2/4M_N^2  \cr \hline
 \rowcolor{celeste} N'  & (1,0)& N' LH & 0 & +3\lambda_N^2/4M_N^2 & +\lambda_N^2/4M_N^2 \cr
  \hline
\rowcolor{verdino} L' & (2,-1/2)& EL'H^* &-\lambda_L^2/2M_L^2 & 0 &0  \cr \hline
\rowcolor{verdino}L_{3/2} & (\bar{2},-3/2) & E(L_{3/2}\epsilon H)   &+\lambda_L^2/2M_L^2 & 0 &0 \cr \hline
\rowcolor{rosa} E'  & (1,1) & E'LH^* & 0 & -\lambda_E^2/4M_E^2 & -\lambda_E^2/4M_E^2 \cr  \hline
\rowcolor{rosa} E^a   & (3,1)&E^a(H^* \tau^a L) & 0 & -3\lambda_E^2/4M_E^2 & +\lambda_E^2/4M_E^2  \cr \hline
\hline\end{array}$$
\caption{\label{tab:listf} List of new leptons that can couple to the SM lepton doublet
$L = (\nu_\mu,l_L)$ or singlet $E=l_R$ 
(with the same gauge quantum numbers as $L'$ and $E'$) and to the Higgs doublet $H=(0,v+h/\sqrt{2}) $ (an SU(2) doublet with $Y=1/2$).
}
\end{small}
\end{table}

\begin{figure}[htb!]
\begin{center}
\includegraphics[scale=0.11]{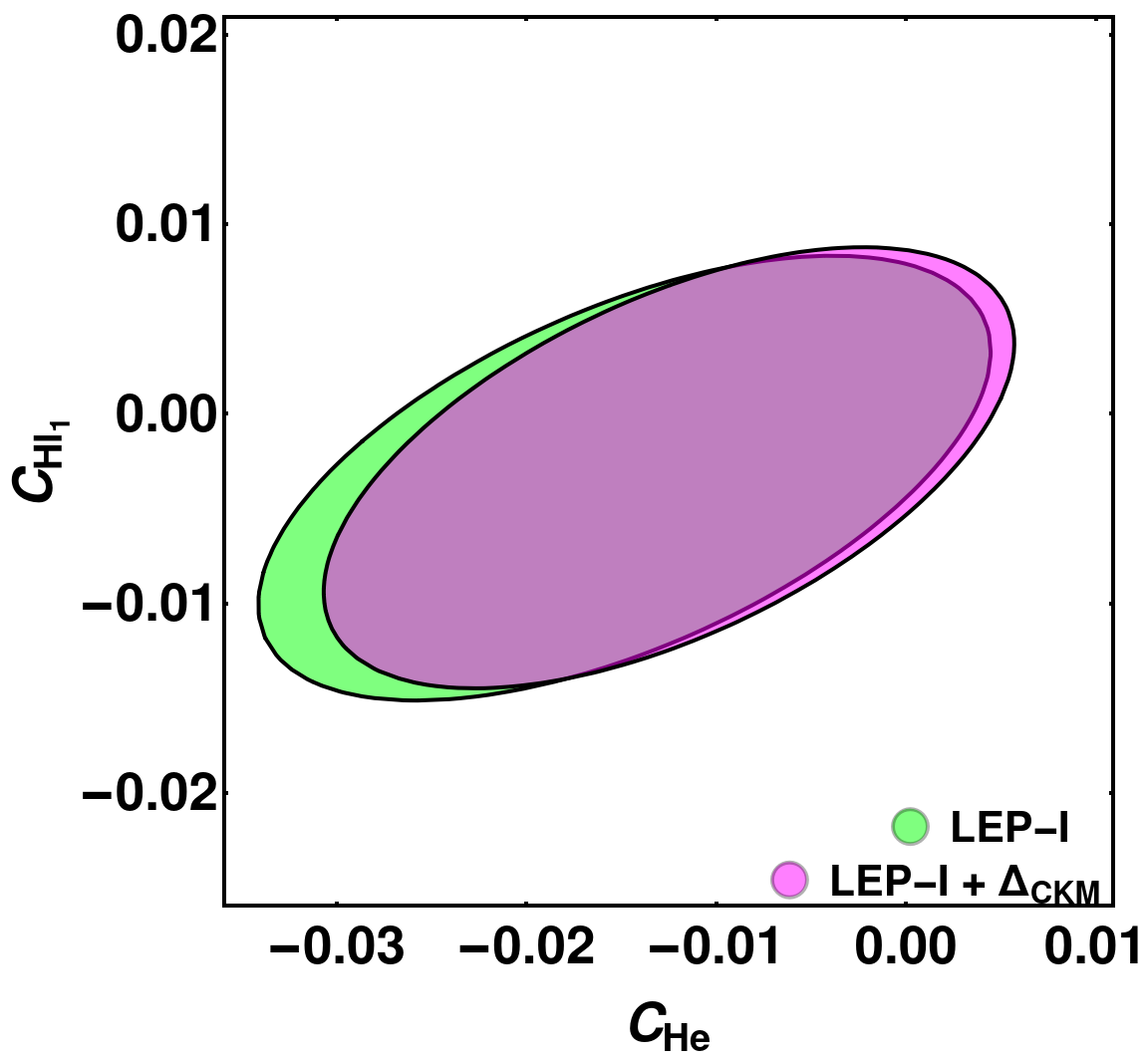}
\includegraphics[scale=0.11]{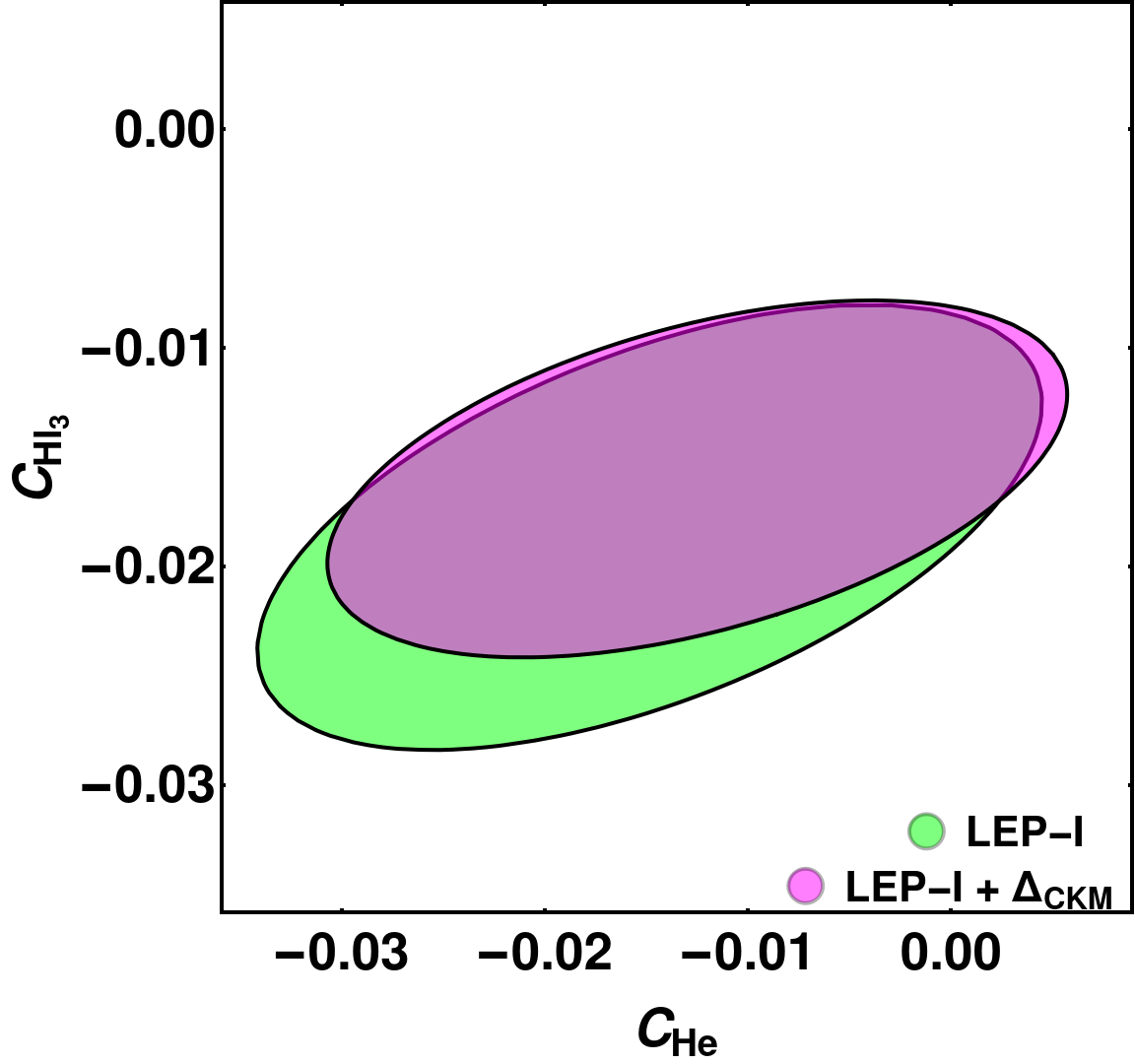}
\includegraphics[scale=0.11]{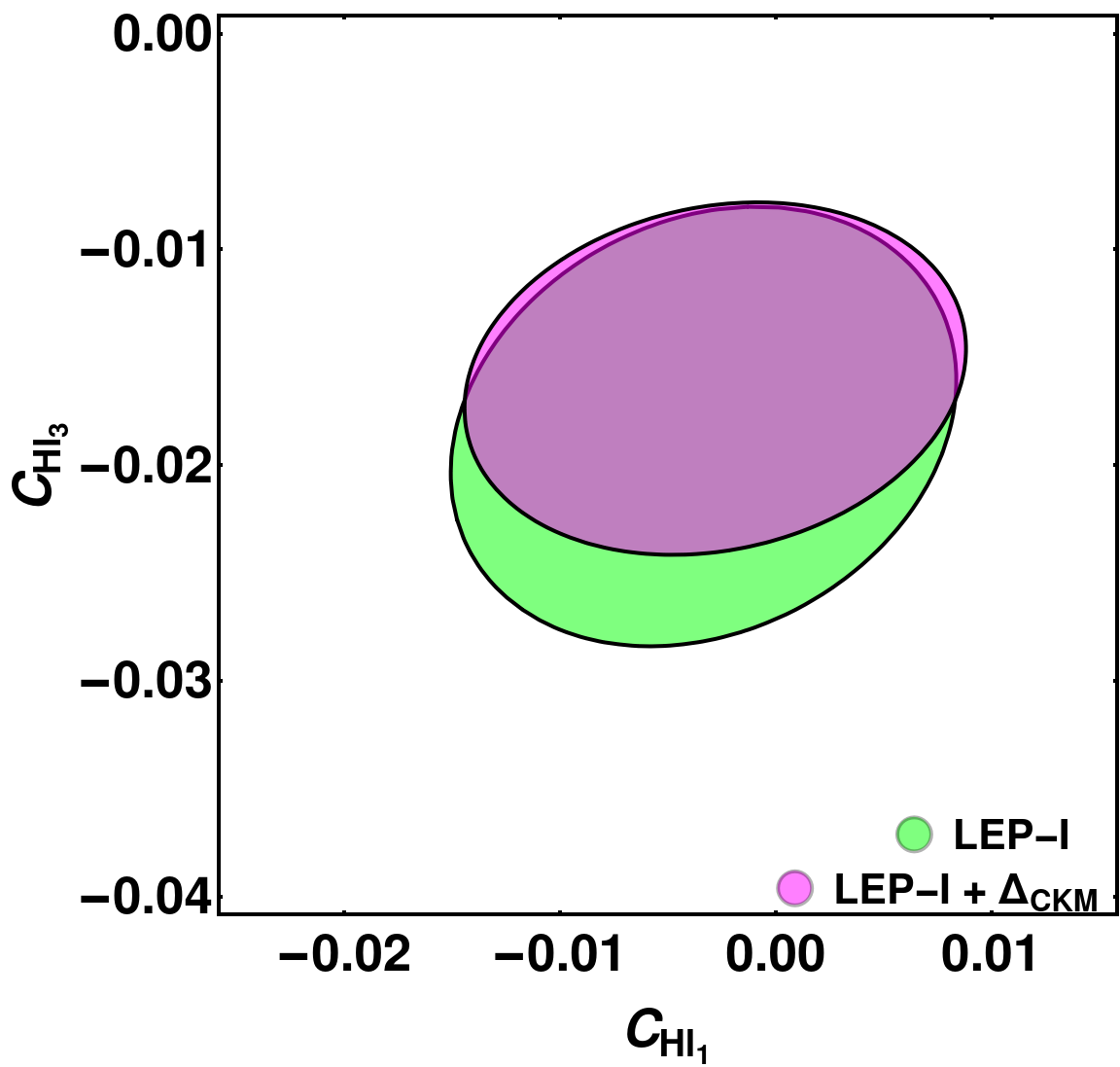}
\end{center}
\caption{2-D Marginalised plot for the WCs impacting the VLLs}
\label{plot-vll}
\end{figure}

Figure~\ref{plot-vll} and Table~\ref{tab:VLF-fit} represents the result of the fit once for the EWPO alone and second for the case including $\Delta_{CKM}$ (matched to SMEFT at 1-loop level) alongwith EWPO. The posterior values of the observables for these two cases can be read off from Table~\ref{table:VLL-post}. We can see that even without including the $\Delta_{CKM}$ constraint, we can satisfy both $M_W$ and $\Delta_{CKM}$ within 2$\sigma$. $A_l^{SLD}$ improves to $1.56\sigma$ while $A_B^{FB}$ deteriorates to $-2.18\sigma$. Inclusion of the $\Delta_{CKM}$ constraint improves the posterior value of $\Delta_{CKM}$ only marginally, while pushing the $M_W$ value beyond 2$\sigma$.
\begin{table}[htb]
    \centering
\begin{small}    
   { \begin{tabular}{|c|c|rrr|c|rrr|}
 \hline
 WC & B.F(\scriptsize{EWPO}) & \multicolumn{3}{c|}{Correlation} & B.F(\scriptsize{EWPO+CKM}) & \multicolumn{3}{c|}{Correlation} \\\hline
 & \multicolumn{4}{c|}{\small{($\chi^2_{\rm fit}$/$\chi^2_{\rm SM}=20.53/40.08$)}} & \multicolumn{4}{c|}{\small{($\chi^2_{\rm fit}$/$\chi^2_{\rm SM}=21.32/44.67$)}} \\
 \hline 
$C_{He}$ & $-0.0148 \pm 0.0075 $ & $1.00$ & & & $-0.0125 \pm 0.0073 $ & $1.00$ & &\\ 
$C_{Hl_1}$ & $ -0.0037\pm 0.0043 $ & $0.57$ & $1.00$ & &$ -0.0027\pm 0.0045 $ & $0.56$ & $1.00$ &  \\ 
$C_{Hl_3}$ & $-0.0184\pm 0.0039$ & $0.55$ & $0.21$ & $1.00$&  $-0.0158\pm 0.0027$ & $0.47$ & $0.17$ & $1.00$ \\ \hline
    \end{tabular}}
    \caption{Global fit of the WCs affecting VLLs at tree-level to EWPO and $\Delta_{CKM}$.}
    \label{tab:VLF-fit}
\end{small}    
\end{table}
\begin{table}[htb!]
\begin{center}
\includegraphics[scale=0.28]{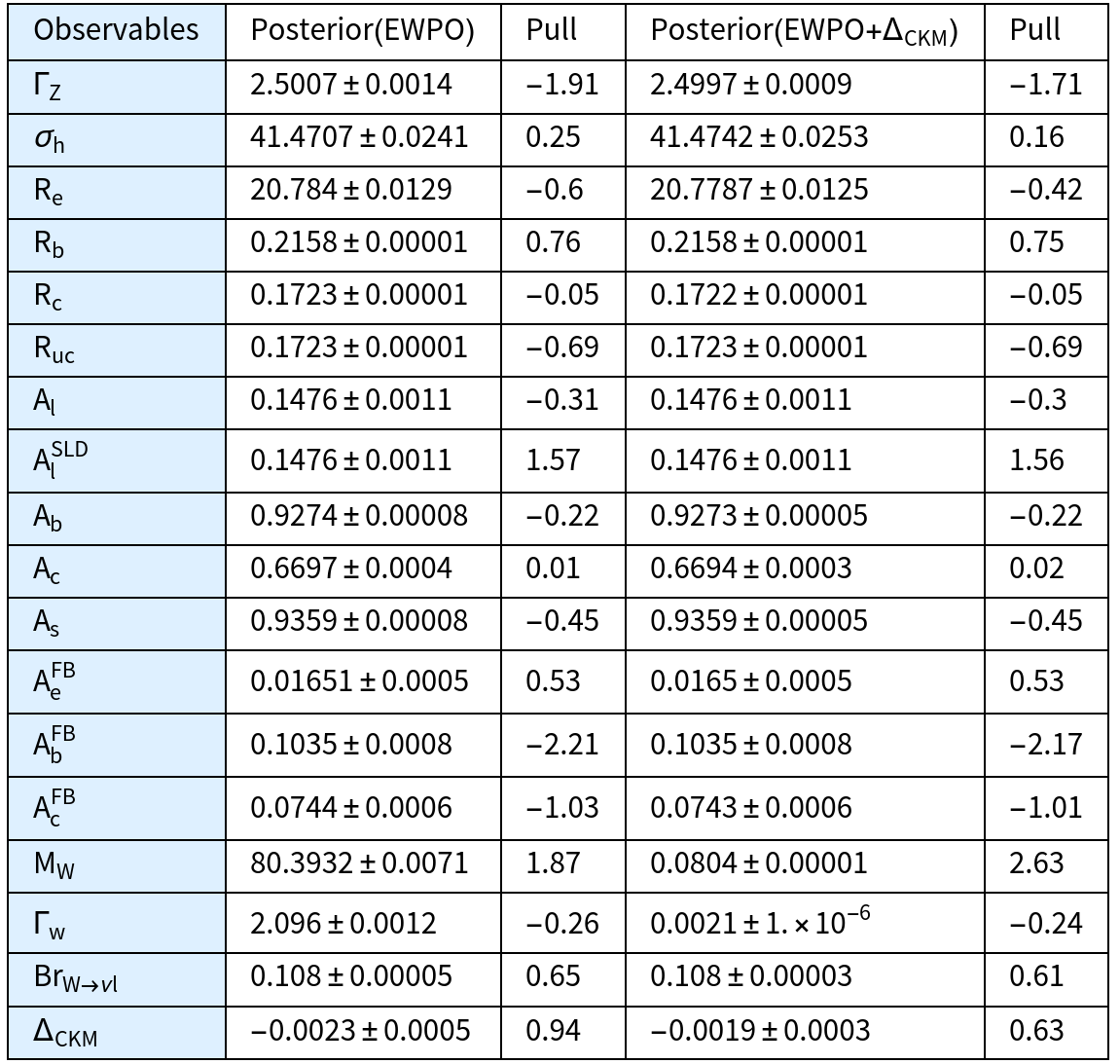}
\caption{Posterior results and Pull for the VLL inspired study, once just with EWPO observables and once including the $\Delta_{CKM}$ constraint.}
\label{table:VLL-post}
\end{center}
\end{table} 
We can put limits on the ratio of the coupling and mass for these scenarios. For example, if we consider only $N'$, we get a limit of ${v \lambda_N}/{M_N} < 0.074$, where $v$ is the vev. Similarly for $E'$, we get a limit of ${v \lambda_E}/{M_E} < 0.05$.
If we look at $L' \oplus E'$, the bounds come out to be ${v \lambda_L}/{M_L} < 0.045$ and ${v \lambda_E}/{M_E} < 0.066$.

\subsection{EWPO and $\Delta_{CKM}$: 8 parameter fit}
Though, at leading order, EWPO depend on ten SMEFT operators, only eight among these can be constrained using the electroweak data.  Some of the earlier literature~\cite{deBlas:2022hdk,Falkowski:2014tna} redefined the combinations of the corresponding WCs by rotating away the two purely bosonic contributions ($C_{HD}$ and $C_{HWB}$) to get eight independent parameters which are finally constrained. When analyzed in that fashion, we showed that our results are in well agreement with those in the literature. We also found out that same constraints can also be achieved by setting those two purely bosonic WCs just to zero. Slight variation of results in both our analyses do arise compared to previous analyses because of the choice of the input scheme and inclusion of theoretical uncertanities. Phenomenologically, there can be frameworks where these bosonic operators are highly suppressed and so this procedure can direct provide limits for such kind of scenarios.

\begin{figure}
\begin{center}
\includegraphics[width=15cm,height=20cm]{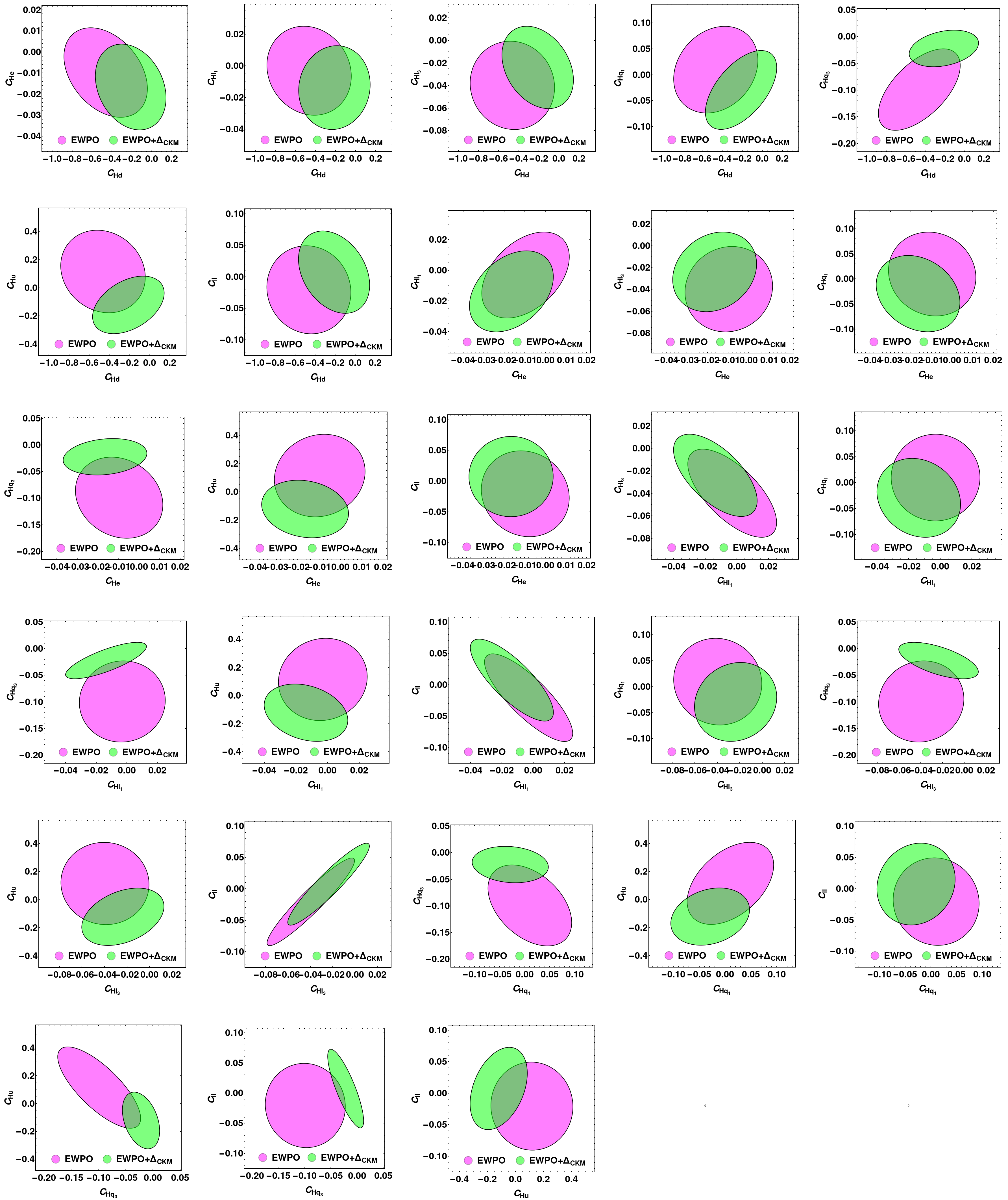}
\end{center}
\caption{2-D marginalised plot for the fit with 8 WCs}
\label{fig:8pplot}
\end{figure}

The results of these analyses are presented in Table~\ref{tab:EWPO:8WC}. The second column of the table shows the best fit and deviation for the WCs given in the first column. Deviations are computed using the marginalization procedure discussed in the section~\ref{sec:ewf}.
We also tabulate the correlations between the coefficients in the last column. Among them, $C_{Hl_3}$ and $C_{ll}$ are highly correlated with a value of $0.94$. The same for $C_{Hl_1}$ and $C_{ll}$ is $0.81$ and for $C_{Hl_1}$ and $C_{Hl_3}$ it is $-0.67$. From the shifts in the gauge coupling shown in Table~\ref{tab:coupling-shifts},  one can expect very small correlations among quark WCs and leptonic one $\sim \mathcal{O}(0.01)$. The value of the SMEFT $\chi^2$ after the fit comes out to be $3.24$ compared the SM $\chi^2$ of $40.08$, implying a very good quality of the fit. The 2-D marginalized distributions of the WCs are shown in Figure~\ref{fig:8pplot} in purple colour.

Posterior values of the EWPO observables show very good agreement with the experimental data with all of them within $1\sigma$ as shown in Table~\ref{8WC-post}. Notable improvements among them are $M_W$ at $-0.2\sigma$, $A_b^{FB}$ at $-0.19\sigma$ and $A_l^{SLD}$ at 0.84$\sigma$. $\Delta_{CKM}$ however becomes worse as the pull with the EWPO fitted parameters increases its discrepancy to $-2.7\sigma$.

\begin{table}[!htb]
\begin{small}
    \centering
    \begin{tabular}{|c|c|rrrrrrrr|}
 \hline
WC & B.F(\scriptsize{EWPO}) & \multicolumn{8}{c|}{Correlation } \\ 
 \hline 
 & \multicolumn{9}{c|}{\scriptsize{($\chi^2_{\rm fit}$/$\chi^2_{\rm SM}=3.24/40.08$)}} \\
 \hline
$C_{Hd}$   &-0.4646 $\pm $ 0.1715 &  1.00 &  & & & & & &  \\ 
$C_{He}$   & -0.0099$\pm $ 0.0085 & -0.36 & 1.00 & & & & &  & \\ 
$C_{Hl_1}$ & -0.0031$\pm $ 0.011  & -0.12 & 0.42 & 1.00 & & & & &  \\
$C_{Hl_3}$ & -0.0398$\pm  $0.0159 & -0.05 & 0.08 &-0.67 & 1.00 &  &  &  &    \\
$C_{Hq_1}$ & 0.0101$\pm $ 0.0341  &  0.21 &-0.08 & 0.01 &-0.04 & 1.00 & &  &  \\
$C_{Hq_3}$ & -0.0989$\pm $ 0.0310 &  0.59 &-0.17 & 0.03 & 0.06 &-0.42 & 1.00 &  &  \\
$C_{Hu}$   &  0.1162$\pm $ 0.1195 & -0.14 & 0.10 & 0.07 &-0.03 & 0.44 &-0.76 & 1.00&   \\
$C_{ll}$   & -0.0204$\pm $ 0.0284 & -0.06 &-0.10 &-0.81 & 0.94 &-0.04 &-0.03 & -0.01 & 1.00 \\ \hline
    \end{tabular}
    \caption{Global fit to EWPO for the case with 8 WCs.}
    \label{tab:EWPO:8WC}
\end{small}
\end{table}

\begin{table}[!htb]
\begin{small}
    \centering
    \begin{tabular}{|c|c|rrrrrrrr|}
 \hline
WC & B.F(\scriptsize{EWPO+$\Delta_{CKM}$}) & \multicolumn{8}{c|}{Correlation } \\ 
 \hline 
 & \multicolumn{9}{c|}{\scriptsize{($\chi^2_{\rm fit}$/$\chi^2_{\rm SM}=10.88/44.67$)}} \\
 \hline
$C_{Hd}$   & -0.2125$\pm $ 0.1453  & 1.00 & & & & & & &  \\ 
$C_{He}$   & -0.0166$\pm $ 0.0083  &-0.24 & 1.00 &  &  &  &  &  &  \\ 
$C_{Hl_1}$ & -0.0136$\pm $ 0.0107  & 0.08 & 0.35 & 1.00 &  &  &  &  &  \\
$C_{Hl_3}$ & -0.0238$\pm $ 0.0148  &-0.31 & 0.19 &-0.63 & 1.00 &  &  &  &  \\
$C_{Hq_1}$ & -0.0291$\pm $ 0.0311  & 0.55 &-0.23 &-0.16 & 0.12 & 1.00 &  &  &\\
$C_{Hq_3}$ & -0.0221$\pm $ 0.0139  & 0.30 & 0.18 & 0.78 &-0.64 &-0.13 & 1.00 & & \\
$C_{Hu}$   & -0.1206$\pm $ 0.0832  & 0.40 &-0.15 &-0.27 & 0.37 & 0.23 &-0.37 & 1.00 &\\
$C_{ll}$   & 0.0076$\pm $ 0.0265 &-0.32 & 0.01 &-0.79 & 0.93 & 0.13 &-0.83 &0.38 &1.00 \\ \hline
\end{tabular}
    \caption{Global fit to EWPO and $\Delta_{CKM}$ for the case with 8 WCs.}
    \label{tab:EWPO-CKM:8WC}
\end{small}    
\end{table}

\begin{table}[!htb]
\centering
\includegraphics[scale=0.3]{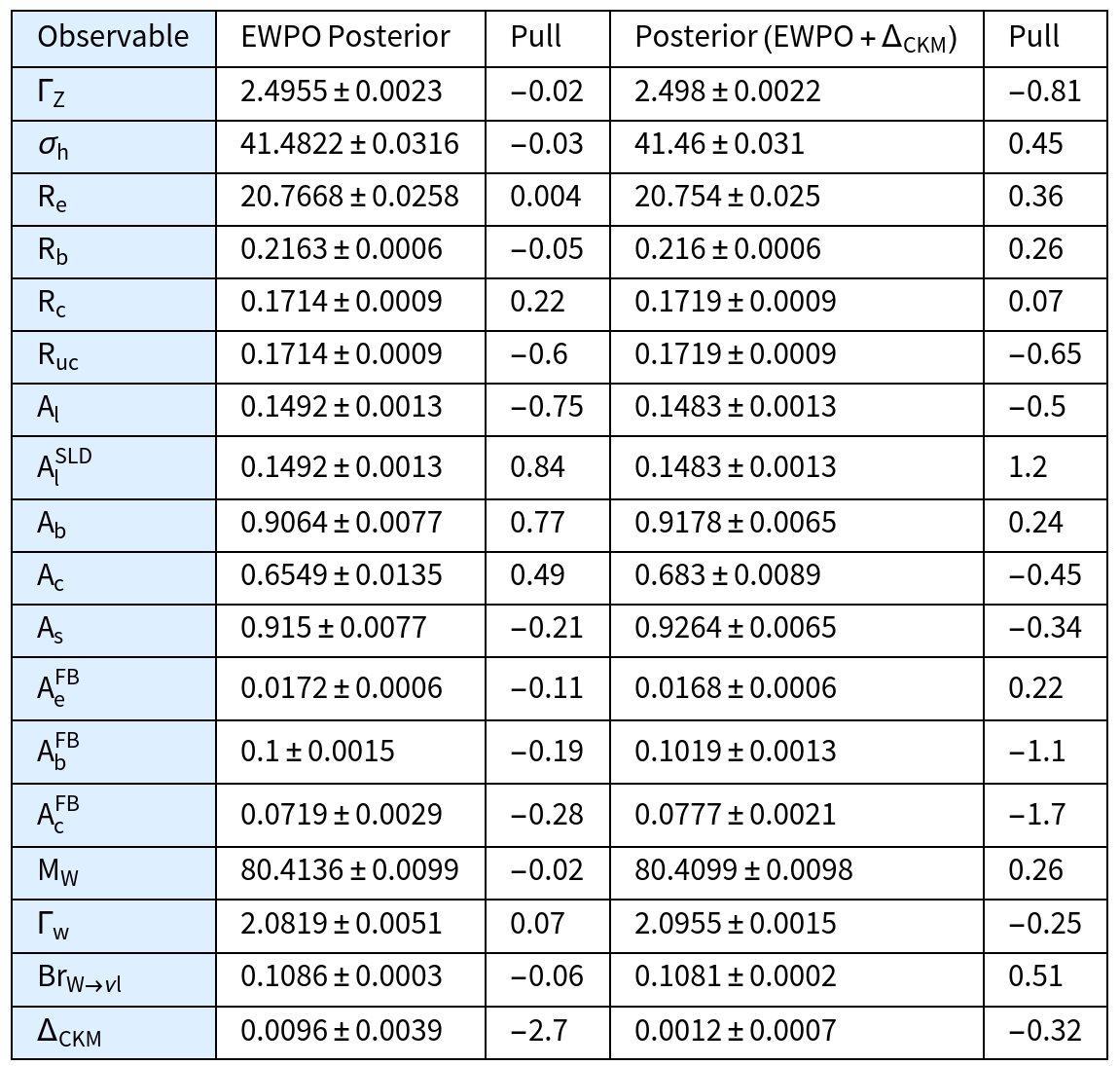}
\caption{Posterior Value and Pulls for the EWPO observables with 8 WCs, once without $\Delta_{CKM}$ and once with $\Delta_{CKM}$}
\label{8WC-post}
\end{table} 

Inclusion of the CKM anomaly ($\Delta_{CKM}$ matched to SMEFT at 1-loop level), along with EWPO, shifts the best-fit values of the WCs, constraining them better. The results of the fit can be read off from Table~\ref{tab:EWPO-CKM:8WC} and the 2-D marginalized distributions of the WCs are given in Figure~\ref{fig:8pplot} in green colour.
$\Delta_{CKM}$ has the largest dependence on $C_{Hl_1}$, $C_{ll}$, $C_{Hl_3}$ and $C_{Hq_3}$; therefore there is significant change in the correlations involving these WCs. $C_{Hq_3}$ and $C_{ll}$ become highly correlated with a value of $-0.83$. As $\Delta_{CKM}$ depends only on left chiral operators, the correlation of their WCs to that of the right chiral ones decrease. $\Delta_{CKM}$ also introduces a high correlation between the WCs corresponding to leptonic and quark operators, e.g., the correlation factor for $C_{Hq_3}$ with $C_{Hl_1}$ and $C_{Hl_3}$ now stands at 0.78$\sigma$ and $-0.64\sigma$ respectively.

On the other hand, the inclusion of $\Delta_{CKM}$ in the fit observable reduces its discrepancy to a mere $-0.32\sigma$. However, discrepancy in $A_{c}^{FB}$, $A_{b}^{FB}$ and $A_l^{SLD}$ goes beyond 1$\sigma$. The other observables also show good agreement with the experimental observations.

\subsection{Including LEP-II data to EWPO and $\Delta_{CKM}$: 10 parameter fit} 
As discussed in the previous subsection, EWPO can only constrain eight of the 10 WCs. This can be overcome by including LEP-II data. During the second run of the LEP, sufficient energy was reached for the production of a pair of on-shell $W$ bosons, dominant contribution coming from a $t$-channel neutrino exchange. This provides an opportunity to test four-fermionic final states at the collider. Precise measurements of cross-sections and angular distributions at various energies enhance the list of precision observables. Ref.~\cite{Han:2004az,Berthier:2016tkq,Ellis:2018gqa} also explored this in the context of SMEFT and showed that the presence of the subdominant $s$-channel production of a pair of $W$ bosons through a $\gamma$ or $Z$ exchange lifts up the flat directions which were there in the case of Z-pole EWPO. In addition to the 10 WCs already affecting the EWPO, introduction of LEP-II data brings dependence on the anomalous triple gauge operator [$\mathcal{O_W} \equiv {W}_{\mu}^{ \nu}{W}_{\nu}^{ \rho} {W}^{\mu}_{\rho}$] with the WC $C_W$ (in the Warsaw basis) in the set of operators contributing at tree-level. As this is not part of EWPO or $\Delta_{CKM}$ we set $C_W=0$ for our analyses.

\begin{figure}[!htb]
\centering
\includegraphics[scale=0.14]{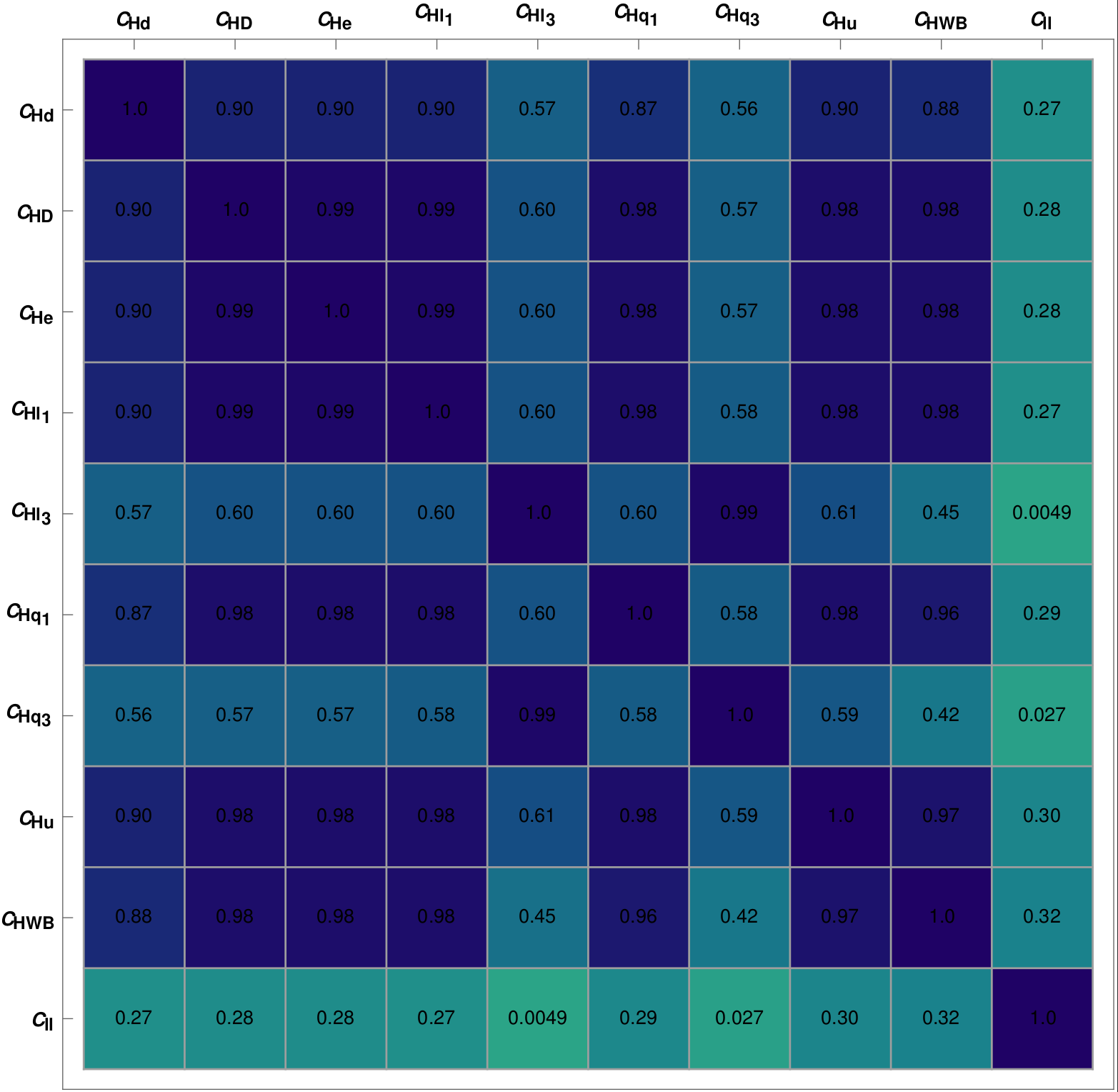}~
\includegraphics[scale=0.14]{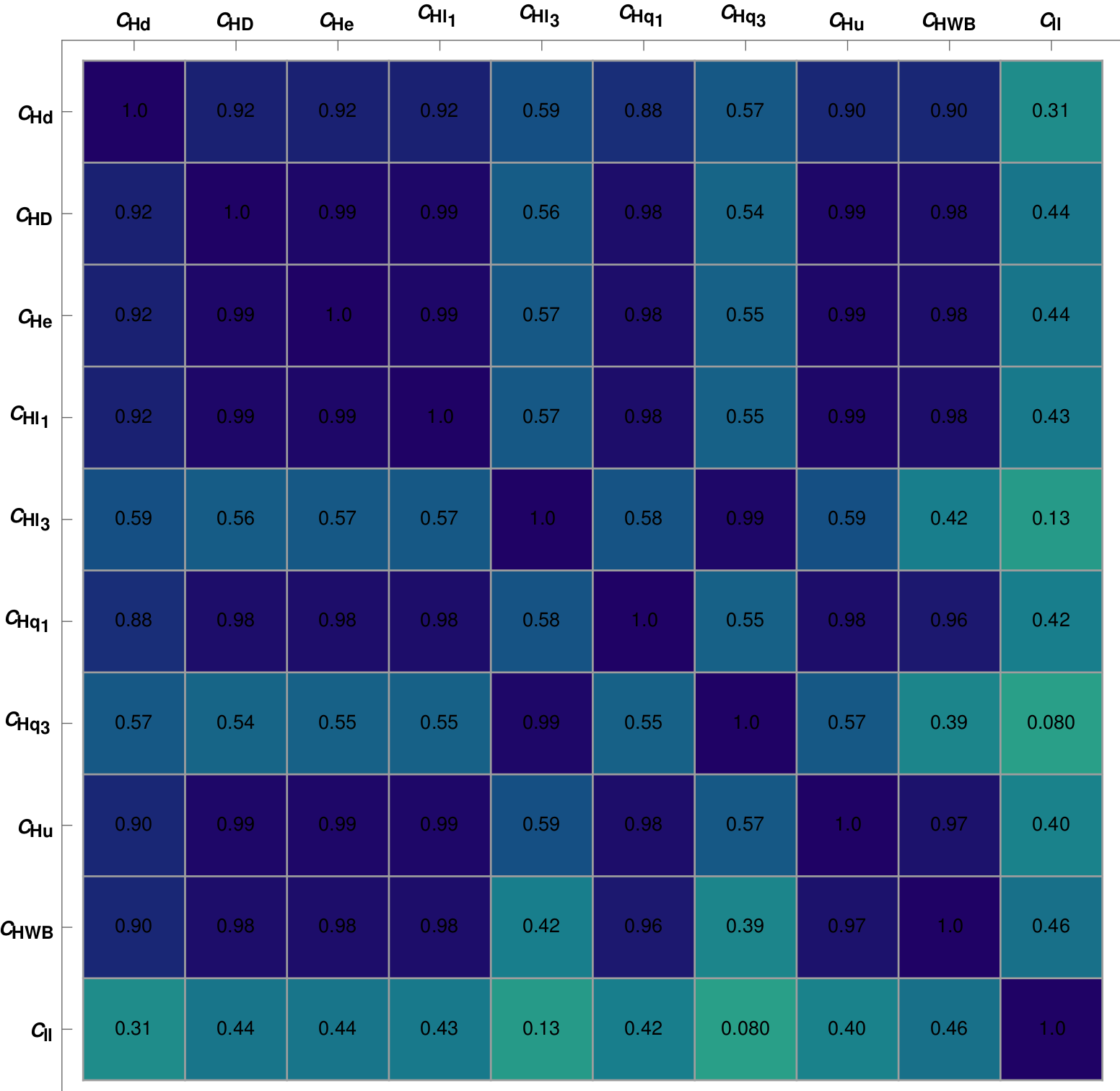}
\\
(a) \hspace{7cm} (b)
\caption{Density plot of correlation matrix after including LEP-II data: (a) EWPO (b) EWPO+$\Delta_{CKM}$.}
\label{fig:cor10wc}
\end{figure}

Including the LEP-II data increases the $\chi^2$ to 121 due to the increase in number of observables. The results of the fit without including $\Delta_{CKM}$ are shown in the Table~\ref{tab:EWPO:10WC}. The minimum value of $\chi^2$ now decreases to $83.8$. But, the best fit values of $C_{HD}$, $C_{HWB}$ and $C_{He}$ are slightly on the larger side. The correlations among the WCs change drastically due to the presence of the new LEP-II observables. Correlation between $C_{Hl_3}$ and $C_{ll}$ becomes almost negligible while many pairs of WCs become highly correlated. To show the strength of correlation we also plot color density matrix plot in Figure~\ref{fig:cor10wc}(a). Highest correlation is represented by darkish blue colour and gradually turns towards green as strength of the correlation decreases. It can be easily observed that many of them are dark blue with correlation $\sim 0.9$. The correlation of all WCs with $C_{ll}$ is small $\sim 0.3$ and almost no correlation with $C_{Hl_3}$ and $C_{Hq_3}$. The correlation of $C_{Hl_3}$ and $C_{Hq_3}$ is almost 1 whereas both of them are less correlated with others. The posterior results of the fit for the precision observables are shown in  Table~\ref{tab:10WC-post}. The discrepancy in all of these observables are within $1\sigma$ with the exception of $\Delta_{CKM}$ where the pull for is $-2.7\sigma$. 

Incorporation of $\Delta_{CKM}$ (matched to SMEFT at 1-loop level) as an observable in the fit makes it more constraining as shown in Table~\ref{tab:EWPO-CKM:10WC}. The absolute best-fit values of the WCs lie well within 1 with $C_{ll}$ becoming very small. There is no significant change found in $1\sigma$ deviations of WCs and the correlations among the pairs of WCs also do not show much change. The density plot for correlation matrix assuming magnitude of elements is also shown in Figure~\ref{fig:cor10wc}(b).  

Posteriors results and pull of the EWPO and $\Delta_{CKM}$ are collected in the fourth column and fifth column respectively of Table~\ref{tab:10WC-post}. All the pulls are again within $1\sigma$. As expected, pull for $\Delta_{CKM}$ improves to $-0.36\sigma$. The other observables continue to remain well within 1$\sigma$ making the fit highly competitive with respect to the experimental data. 

\begin{table*}[]
\begin{scriptsize}
    \centering
    \begin{tabular}{|c|c|rrrrrrrrrr|}
 \hline
 & Result & \multicolumn{10}{c|}{Correlation } \\ 
 \hline 
 & \multicolumn{11}{c|}{\scriptsize{($\chi^2_{\rm Fit}$/$\chi^2_{\rm SM}=83.82/121.79$)}} \\
 \hline
$C_{Hd}$   & -0.0841$\pm $ 0.3937 & 1.00 & & & & & & & & &  \\ 
$C_{HD}$   & -2.1541$\pm $ 2.2945 & -0.90 & 1.00 & & & & & & & &\\ 
$C_{He}$ & 1.0669$\pm $ 1.1474 & 0.90 & -0.99 & 1.00 & & & & & & & \\
$C_{Hl_1}$ & 0.5383$\pm $ 0.5713 & 0.90 & -0.99 & 0.99 & 1.00 & & & & & &\\
$C_{Hl_3}$ & 0.01423$\pm $ 0.4149 & 0.58 & -0.60 & 0.60 & 0.61 & 1.00 & & & & &\\
$C_{Hq_1}$ & -0.1717$\pm $ 0.1912 & -0.87 & 0.98 & -0.98 & -0.98 & -0.60 & 1.00 & & & & \\
$C_{Hq_3}$   & -0.0346$\pm $ 0.41 & 0.57 & -0.58 & 0.58 & 0.58 & 0.99 & -0.58 & 1.00 & & & \\
$C_{Hu}$   & -0.6179$\pm $ 0.7532 & -0.90 & 0.99 & -0.99 & -0.99 & -0.61 & 0.98 & -0.59 & 1.00 & & \\ 
$C_{HWB}$   & 1.0071$\pm $ 0.9923 & 0.88 & -0.98 & 0.98 & 0.98 & 0.46 & -0.97 & 0.43 & -0.97 & 1.00 & \\
$C_{ll}$   & -0.0295$\pm $ 0.0273 & 0.27 & -0.29 & 0.29 & 0.27 & 0.005 & -0.30 & -0.028 & -0.30 & 0.32 & 1.00\\
\hline

\end{tabular}
    \caption{Best fit with $1\sigma$ deviation of WCs after including LEP-II data along with EWPO. Correlation matrix of coefficients are also shown in the third column.}
    \label{tab:EWPO:10WC}
\end{scriptsize}    
\end{table*}

\begin{table*}[]
\begin{scriptsize}
    \centering
    \begin{tabular}{|c|c|rrrrrrrrrr|}
 \hline
 & Result & \multicolumn{10}{c|}{Correlation } \\ 
 \hline 
 & \multicolumn{11}{c|}{\scriptsize{($\chi^2_{\rm Fit}$/$\chi^2_{\rm SM}=90.93/126.37$)}} \\
 \hline
$C_{Hd}$   & -0.1266$\pm $ 0.3934 & 1.00 & & &  &  &  &  &  &  &   \\ 
$C_{HD}$   & -0.5455$\pm$ 2.2138 & -0.93 & 1.00 &  &  &  &  &  &  &  & \\ 
$C_{He}$ & 0.2564$\pm $ 1.1066 & 0.93 & -0.99 & 1.00 &  &  &  &  &  &  &  \\
$C_{Hl_1}$ & 0.125$\pm $ 0.5498 & 0.93 & -0.99 & 0.99 & 1.00 & & & & & & \\
$C_{Hl_3}$ & -0.3029$\pm $ 0.3974 & 0.59 & -0.57 & 0.57 & 0.58 & 1.00 & & & & & \\
$C_{Hq_1}$ & -0.07295$\pm $ 0.1876 & -0.88 & 0.98 & -0.99 & -0.99 & -0.58 & 1.00 &  & & & \\
$C_{Hq_3}$   & -0.2994$\pm $ 0.403 & 0.57 & -0.55 & 0.55 & 0.56 & 0.99 & -0.56 & 1.00 & & &\\
$C_{Hu}$   & -0.2947$\pm $ 0.7434 & -0.90 & 0.99 & -0.99 & -0.99 & -0.60 & 0.98 & -0.58 & 1.00 & &\\ 
$C_{HWB}$   & 0.4054$\pm $ 0.9663 & 0.90 & -0.98 & 0.98 & 0.98 & 0.42 & -0.97 & 0.39 & -0.97 & 1.00 & \\
$C_{ll}$   & -0.00004$\pm $ 0.0249 & 0.31 & -0.44 & 0.44 & 0.43 & 0.14 & -0.42 & 0.08 & -0.41 & 0.47 & 1.00\\
\hline
\end{tabular}
    \caption{Best fit with $1\sigma$ deviation of WCs after including LEP-II data along with EWPO and $\Delta_{CKM}$. Correlation matrix of coefficients are also shown in the third column.}
    \label{tab:EWPO-CKM:10WC}
\end{scriptsize}    
\end{table*}


\begin{table}[!htb]
\centering
\includegraphics[scale=0.6]{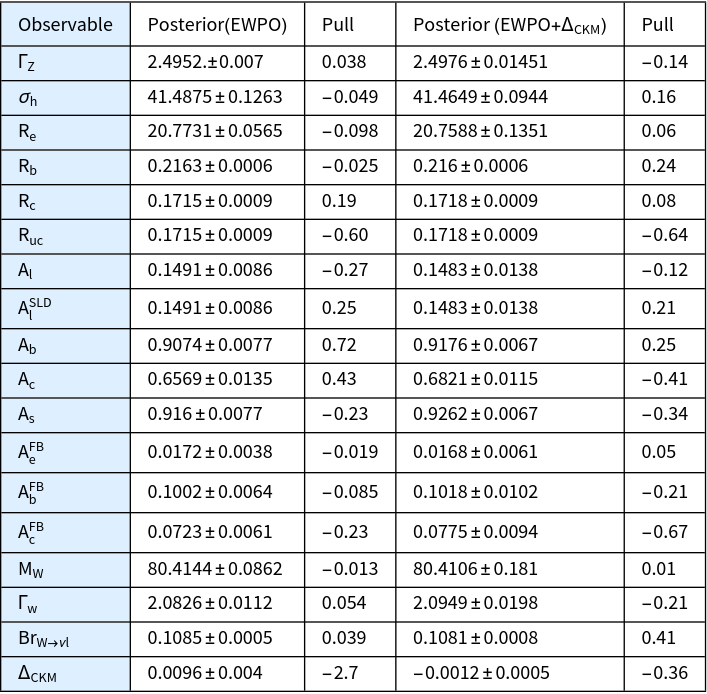}
\caption{Observable's posterior values for the all-parameter fit with 10 WCs }
\label{tab:10WC-post}
\end{table} 

\section{Summary and conclusions}
\label{sec:outlook}
Electroweak precision measurements place some of the most stringent constraints on all New Physics extensions of Standard Model. We explore dimension-6 operator subsets of the SMEFT in the context of Electroweak Precision data and the recent $M_W$ anomaly with the latter hinting at the presence of BSM physics. In addition to the precision data, we also analyze the status of the Cabibbo anomaly within SMEFT using the most recent data. To implement it in a more precise way, we compute the contributions to the beta decay and muon decay using the SMEFT operators that enters in the LEFT at one loop matching. The results of this model-independent study are, then, used to investigate a few specific UV-complete BSM scenarios that are inspired by various anomalies in the experimental data.

To begin with, we parameterize the impact of BSM heavy states in oblique parameters $S$, $T$ and $V$ and compute the posterior fit of the EWPO and $\Delta_{CKM}$. The results are impressive as pulls for almost all the observables (except $A_l^{SLD}$ and $A_b^{FB}$) are within $1\sigma$. Another point to note is that even without including $\Delta_{CKM}$ into the fit the pull for Cabibbo anomaly comes down to $\sim -0.52$. On including $\Delta_{CKM}$, the pull improves to $\sim -0.27$.  In the next study, we choose the SMEFT operators $C_{Hl_3}$ and $C_{ll}$ as the ones which dictates the UV BSM physics as both of them affect the $M_W$, $G_F$ and $\Delta_{CKM}$ at tree level. We analyze this case for EWPO as well as the EWPO+$\Delta_{CKM}$ ($\Delta_{CKM}$ matched to SMEFT at 1-loop level) combination. Although they can explain the Cabibbo anomaly very well, they fare very badly with $M_W$ and $A_b^{FB}$. 

Since $M_W$ has the highest pull away from the SM among the observables of our interest, we next considered the BSM case which consists of the four  operators that affect the $M_W$ at tree level ($C_{HD}$, $C_{HWB}$, $C_{Hl_3}$ and $C_{ll}$). The minimum value of $\chi^2$ comes down to $\sim ~11$ from $\sim ~40$ and the fit fares very well with $M_W$ with discrepancy well within 1$\sigma$.  Bosonic operators $C_{HD}$ and $C_{HWB}$ are highly correlated in this case. Inclusion of $\Delta_{CKM}$ (matched to SMEFT at 1-loop level) shifts the best fit towards slightly higher values while simultaneously shrinking the allowed 2$\sigma$ ranges. Discrepancies in $A_l^{SLD}$ and $A_b^{FB}$ continues to remain towards the higher side.

We then choose the set of operators motivated from Vectorlike lepton (VLL) model. Integrating out heavy degrees of freedom, at tree-level, generates the operators $C_{He},C_{Hl_1}$ and $ C_{Hl_3}$. The allowed range for these WCs, after constraining using EWPO and $\Delta_{CKM}$, turn out to be $\mathcal{O}(0.01)$. Inclusion of $\Delta_{CKM}$ (matched to SMEFT at 1-loop level) shifts allowed ranges to slightly larger values. We also explore a few minimal VLL frameworks and set the constraints over ratios of VLF Yukawas and the scale of the model. The posterior fit of the observables show that the pull for $\Gamma_Z$ becomes worse with a discrepancy over $1.5\sigma$. New posterior value of $M_W$ is improved in comparison to SM, but the pull remains at $\sim 1.9 \sigma$ and $\sim 2.4 \sigma$ for the fits without and with $\Delta_{CKM}$ respectively. Thus we can conclude that minimal VLL frameworks are severely constrained by the precision measurements and slight tension exists with recent measurement of the $M_W$ mass and $A_b^{FB}$. 

Considering all the dimension-6 SMEFT operators, that appear in the EWPO at the leading order, only eight of the ten WCs, that appear at the tree-level, can be constrained by the EWPO. We compute their best fit, first with only EWPO and then including $\Delta_{CKM}$ (matched to SMEFT at 1-loop level). For the case of only EWPO, the best fit $\chi^2$ becomes $\sim 3$ compared to a SM $\chi^2$ value of $40$. On the other hand, the inclusion of $\Delta_{CKM}$ makes the best fit $\chi^2$ $\sim 10.8$ compared to a SM $\chi^2$ of $44$.  We also analyzed 2D correlations of the pairs of WCs with the highest correlation being between $C_{Hl_3}$ and $C_{ll}$. We find that SMEFT at the leading order brings the pulls for all the Electroweak observables within $1\sigma$. Recent $M_W$ measurement also fares exceptionally well with the pull coming down to $\sim -0.03 \sigma$. However, the EWPO fit worsens the $\Delta_{CKM}$ compared to the SM as the discrepancy increases to $-2.7 \sigma$. When the $\Delta_{CKM}$ constraint is also taken into the account in the fit, discrepancies in $A_l^{SLD}$, $A_b^{FB}$ and $A_c^{FB}$ go beyond 1$\sigma$ while improving the agreement with $\Delta_{CKM}$ at $-0.3 \sigma$. Overall, we can conclude that even after including the $\Delta_{CKM}$ into the fit, the pulls for all the observables are within $2\sigma$ with most them within $1\sigma$.

Usage of the LEP-II data lifts the two blind directions which are otherwise present for the EWPO. These new observables correspond to the pair production of $W$'s, which subsequently leads to four fermion final states, resulting in a number of angular observables at various energies. For the case without including the $\Delta_{CKM}$ constraint, limits on the purely bosonic WCs $C_{HD}$ and $C_{HWB}$ are rather weak and their best-fit comes out to be slightly on the higher side. The correlations among the WCs are also generally on the higher side. The pulls, however, on the posterior values of the precision observables are very small showing excellent agreement with the experimental data. Like the previous 8 parameter fit, this also worsens the $\Delta_{CKM}$ by increasing the pull to $-2.7\sigma$. The inclusion of $\Delta_{CKM}$ constraint (matched to SMEFT at 1-loop level) in the fit constraints the WCs further, including that of $C_{HD}$ and $C_{HWB}$ along with simultaneously making their best-fit values smaller. Another peculiar aspect of this fit is that it makes the best-fit value of $C_{ll}$ very small. Correlations among the WCs still continues to be on the higher side and the agreement of the posterior values of the precision observables including $\Delta_{CKM}$ with the experimental data still remains excellent.

Our subsequent aim is to automatise this code with the current observables and make it available for public use very soon. Finally, in order to realise a true global fit, we would also be including the LHC observables in future.


\section*{Acknowledgments}
The authors would like to thank Dr.Nilanjana Kumar for valuable discussions during the initial days of the work. M.T.A. acknowledges the financial support of Department of Science and Technology, Government of India, India (DST) through INSPIRE Faculty grant DST/INSPIRE/04/2019/002507.  K.D. acknowledges Council for Scientific and Industrial Research (CSIR), India for JRF/SRF fellowship with the award letter no. 09/045(1654)/2019-EMR-I. K.D. also acknowledges the research Grant No. CRG/2018/004889 of the SERB,
India, for the help with access to computers and accessories bought under the aegis of this grant. T.S. would like to acknowledge the support from the Dr. D.S. Kothari Postdoctoral fellowship scheme no. F.4-2/2006 (BSR)/PH/20-21/0163.
\section{Appendix}
\label{sec:appendix}
The general correction to $\hat{\sigma}_h^0$ {\it near} the $Z$ pole ($s -M_Z^2 \equiv \Delta$) in the SMEFT is
\bea\label{hadronpole}
\frac{\delta \sigma_h^0}{\sigma_h^0} &\simeq& \frac{\delta \Gamma_{Z \rightarrow \ell \bar{\ell}}}{ \Gamma_{Z \rightarrow \ell \bar{\ell}}}
+  \frac{\delta \Gamma_{Z \rightarrow Had}}{ \Gamma_{Z \rightarrow Had}} - \frac{\delta \omega(M_Z^2)}{\omega(M_Z^2)}- \frac{\delta \omega^\star(M_Z^2)}{\omega^\star(M_Z^2)}.
\eea

where \bea
 \bar{w}(s) &=& s \frac{\bar{\Gamma}_Z}{\bar{M}_Z} \text{ we get : } \delta w(s) = s \left(\frac{(\Gamma_Z)_{SM}}{\hat{m}_Z}\right)\left( \frac{\delta \Gamma_Z}{(\Gamma_Z)_{SM}}\right). \\
 \bar{w}(s) &=&  \bar{\Gamma}_Z \bar{M}_Z \text{ we get : } \delta w(s) =  (\Gamma_Z)_{SM} \hat{m}_Z\left( \frac{\delta \Gamma_Z}{(\Gamma_Z)_{SM}}\right).
\eea

but we note the following simplified expressions.
In the SMEFT $\bar{A}_{f}$ can be written as
\bea
\bar{A}_f = \frac{2 \bar{r}_f}{1 + \bar{r}_f^2},
\eea
where $\bar{r}_f = \frac{\bar{g}^{f}_V}{\bar{g}^{f}_A}$. The redefinition of the $Z$ coupling then leads to a shift of $\bar{A}_f$ such that $\bar{A}_f = (A_{f})_{SM} \left( 1 + \frac{\delta A_{f}}{(A_{f})_{SM}}\right)$ where
\bea
\frac{\delta A_{f}}{(A_{f})_{SM}} = \delta r_f \left( 1 - \frac{2 (r_{f}^2)_{SM}}{1+ (r_{f}^2)_{SM}}\right).
\eea
Here $\delta r_f$ is defined by $r_f = (r_{f})_{SM} \left( 1 + \delta r_{f} \right)$ with $\delta r_{f} =  \delta g^{f}_V/ G^{f}_V - \delta g^{f}_A/ G^{f}_A$. We again use : $(...)_{SM}$ for leading order SM predictions and $G^{f}_{A,V}$ for leading order SM predictions for the couplings.
\vspace{0.5cm}
\bibliographystyle{JHEPCust}
\bibliography{vecB}
\end{document}